\documentclass[preprint2]{aastex} 
\usepackage{verbatim}
\newcommand{\beq}{\begin{equation}}
\newcommand{\eeq}{\end{equation}}

\newcommand{\kms}{\,km~s$^{-1}$ } 
 
\newcommand{\hmpc}{\,$h^{-1}$Mpc }
\newcommand{\hmpcii}{\,$h^{-1}$Mpc}

\newcommand{\qvl}{\,DR7-Sub }

\newcommand{\qvlii}{\,DR7-Sub}

\newcommand{\hgpcii}{\,\ensuremath{h^{-3}\mathrm{Gpc}^3}}
\newcommand{\baf}{\,baryonic acoustic feature }
\newcommand{\bafii}{\,baryonic acoustic feature}
\newcommand{\Epaper}{\,\cite{eisenstein05b} }

\newcommand{\GpaperA}{\,\citet*{gaztanaga08iv} }

\newcommand{\Gpaper}{\,GCH }
\newcommand{\Gpaperii}{\,GCH}

\newcommand{\Kpaper}{\,\cite{kazin09a} }
\newcommand{\Kpaperii}{\,\cite{kazin09a}}
\newcommand{\Kpaperalt}{\,\citealt{kazin09a}}

\newcommand{\avg}[1]{\ensuremath{\langle{#1}\rangle}}

\begin{document}

\title{Regarding the Line-of-Sight \\ Baryonic Acoustic Feature \\ in the Sloan Digital Sky Survey \\ 
and Baryon Oscillation Spectroscopic Survey \\ Luminous Red Galaxy Samples}
\author{Eyal A. Kazin,$^a$\footnote{eyalkazin@gmail.com} \ Michael R. Blanton,$^a$ Rom$\acute{\mathrm{a}}$n Scoccimarro,$^a$  \\
Cameron K. McBride,$^b$ Andreas A. Berlind$^b$   }

\affil{$^a$\footnotesize{Center for Cosmology and Particle Physics, New York University, 4 Washington Pl., New York, NY 10003, USA}}
\affil{$^b$\footnotesize{Department of Physics and Astronomy, Vanderbilt University, 1807 Station B, Nashville, TN 37235, USA}}

\begin{abstract}

We analyze the line-of-sight baryonic acoustic feature in the 
two-point correlation function $\xi$ of the Sloan Digital Sky Survey (SDSS)  
luminous red galaxy (LRG) sample ($0.16<z<0.47$). 
By defining a narrow line-of-sight region, $r_p<5.5$\hmpcii,  
where $r_p$ is the transverse separation component,  
we measure   
a strong excess of clustering at $\sim 110$\hmpcii, 
as previously reported in the literature. 
We also test these results in an alternative coordinate 
system, by defining the line-of-sight as 
 $\theta<3^\circ$, where $\theta$ is the opening angle. 
This clustering excess appears much stronger than the 
feature in the better-measured monopole.
A fiducial $\Lambda$CDM non-linear model in redshift-space 
predicts a much weaker signature. 
We use
realistic mock catalogs 
to model the expected signal 
and noise. 
We find that the line-of-sight 
measurements can be 
explained well by 
our mocks as well 
as by a featureless $\xi=0$. 
We conclude 
that there is no convincing evidence 
that the strong clustering measurement is 
the line-of-sight \bafii. 
We also evaluate how detectable such a signal would be in 
the upcoming  Baryon Oscillation Spectroscopic Survey LRG volume (BOSS). 
Mock LRG catalogs ($z<0.6$) suggest that:
({\it{i}} ) the narrow 
line-of-sight cylinder and cone defined above 
probably will not reveal a detectable acoustic feature in BOSS;
({\it{ii}} ) a clustering measurement as high as that in the current sample
can be ruled out (or confirmed) at a high
confidence level using a BOSS-sized data set; 
and
({\it{iii}} ) an
analysis with wider angular cuts, which provide better signal-to-noise ratios, 
can nevertheless be used 
to compare line-of-sight and transverse distances, and thereby
constrain the expansion rate $H(z)$ and diameter distance $D_\mathrm{A}(z)$. 
\end{abstract}

\keywords{cosmology: observation - distance scale - galaxies: elliptical and lenticular, cD - large scale structure of universe}

{\bf
Submitted to The Astrophysical Journal

April 12th 2010
}

\section{Introduction}

The baryonic acoustic feature 
serves as an important tool for our understanding of the 
evolution of the universe  (\citealt{peebles70a}). 
Originating from 
plasma sound-wave residuals 
that came 
to a near stop at the end of
the baryon drag epoch ($z_{d}\sim 1010$), 
it has the potential to 
serve as a cosmic {\it{standard ruler}},
which, in turn, can help us measure 
the expansion 
of the universe 
(\citealt{hubble31}, \citealt{blake03}, \citealt{seo03}, \citealt{hu03a}, \citealt{linder03a}, \citealt{glazebrook05a}). 

In galaxy clustering measurements, 
this feature is a peak of over-density   
at separations of $s\sim100$\hmpc
in the two point correlation function $\xi$, 
which results in an oscillatory feature 
in 
the 
power spectrum P$(k)$. 

\Epaper were the first to detect 
this feature, using the
clustering of 
$\sim 44,000$ 
luminous red galaxies (LRGs) 
from the Sloan Digital Sky Survey (SDSS;  \citealt{york00a}).
Their measurement 
of the angle-averaged $\xi$, 
which is commonly referred to as the {\it{monopole}}, 
has the power to constrain a 
combination of the Hubble expansion rate 
$H(z)$ and the angular diameter distance
$D_\mathrm{A}(z)$. 

To measure $H$ and 
$D_\mathrm{A}$ separately, 
one would like to probe  
the \baf independently 
along the line-of-sight and 
transverse directions (\citealt{matsubara04}).
Measurements of these potentially promising 
methods are currently 
strongly compromised 
by shot noise and 
sample variance limitations,
due the large scale nature 
of the feature. 
For this reason, 
most studies have focused 
on measuring and analyzing the baryonic 
acoustic feature in the angle 
averaged $\xi$ 
(\citealt{martinez08},
\citealt{cabre09i}, 
\citealt{labini09}, 
\citealt{sanchez09a}, 
\Kpaperalt), 
P($k$)  
(\citealt{cole05a}, 
\citealt{tegmark06a}, 
\citealt{hutsi06a},
\citealt{percival07a}, 
\citealt{percival09b}, 
\citealt{reid09a}) 
and the projected two-point 
function of photo-$z$ samples 
(\citealt{padmanabhan07b}, 
\citealt{blake07})
in the SDSS and and Two 
Degree Field Galaxy Redshift Survey (\citealt{colless03a}) 
galaxy samples. 

\GpaperA (hereafter referred to as \Gpaperii),  
however, 
claim to have 
measured 
the line-of-sight baryonic acoustic 
feature.
Using $\sim 77,000$ LRGs 
from the SDSS Data Release 6 
(DR6, \citealt{adelman08a}) and
$\sim 100,000$ from
DR7 (\citealt{abazajian09a}) they report a 
significant detection
of a feature
at $\pi\sim 110$\hmpc in the line-of-sight direction, 
within a projected distance of $r_p<5.5$\hmpcii.

The 
clustering excess 
they focus on
appears much stronger
than expected from the 
\baf according 
to the concordance $\Lambda$CDM 
model.
The galaxies are observed 
in {\it{redshift-space}}, 
as opposed to real-comoving-space. 
In redshift-space ($z$-space) 
the line-of-sight peak 
in the non-linear correlation function
is expected to smear 
heavily due to 
velocity dispersion.  Furthermore, 
the whole correlation function
should appear negative  
due to the strong squashing effect 
(\citealt{kaiser87}) at scales $s>50$\hmpc
in the line-of-sight direction.

These predictions suggest that if the 
sharp strong positive measurement  
obtained by \Gpaper is the real feature, 
it would require  
a physical explanation.  

Magnification bias 
has been proposed 
to increase clustering at the feature scales. 
This effect results from gravitational 
lensing modifying the spacial distribution 
of high redshift objects (\citealt{turner84a}, \citealt{hui07a} and references within). 
\cite{yoo2009a} and \Gpaper 
examine this effect for the 
redshift-space $\xi$ at 
$z=0.35$. 
Both studies agree that the magnification 
effect is anisotropic having the strongest 
impact on the line-of-sight. 
\Gpaper show that a model 
with magnification performs 
slightly better than without  
($2\sigma$ level).  
We do not include an analysis of magnification bias 
in this study, but show that the 
line-of-sight clustering does agree well 
with a fiducial $\Lambda$CDM model without 
a magnification bias. 

In particular, \cite{miralda09} 
argues, using pair-count statistics (based on data analyzed by \Gpaperii), 
that the clustering excess is not significant, 
and should not be 
regarded as
a detection of the \bafii. 
We concur with that conclusion here.

The purpose of this 
study is to revisit the 
line-of-sight clustering signal 
in the SDSS LRG sample,  
examine its reliability  
and predict the 
signal and its uncertainties 
obtainable in the much larger 
volume and denser
Baryonic Oscillation Spectroscopic Survey sample
(BOSS; \citealt{schlegel09a}). 

We measure the
line-of-sight clustering $\xi$   
at sfcales of $40-200$\hmpc, finding 
results similar to that obtained
by \Gpaperii. 
We predict 
$z$-space (as well as real-space) signals  
and uncertainties 
for SDSS-sized volumes,  
by using 
very realistic light-coned mock galaxy
catalogs which 
are based on fiducial 
 $\Lambda$CDM models.

In \S\ref{datamocks} we briefly 
explain the data and mock catalogs used 
for analysis.
In \S\ref{a2pcf_analysis} we present 
the anisotropic $\xi$ clustering and 
the coordinate systems used throughout the study. 
In \S\ref{dr7dim_analysis} we analyze the line-of-sight 
clustering of \qvlii, 
and in \S\ref{dr7full_analysis} we 
perform a similar analysis on the larger DR7-Full, 
and directly compare results with \Gpaperii.  
We examine the significance of the strong 
line-of-sight clustering signal in  \S\ref{interpretation}  
by applying a Jeffreys scale to 
compare model fits to data performed here 
and in \Gpaperii. 
In \S\ref{boss} we predict the line-of-sight 
measurement expected from the BOSS sample, 
along with a detailed comparison 
of the signal-to-noise ratios of the three volumes 
discussed here. 
In \S\ref{boss2} we vary 
the definition of line-of-sight to 
wider wedges, to show that BOSS may 
be used to disentangle $H(z)$ and $D_\mathrm{A}(z)$.

In the following, all calculations assume
a flat $\Lambda$CDM model. 
When converting data redshifts to comoving 
distances, 
we assume a present day matter density $\Omega_{M0}$=$0.25$,
and define $H_{0}=100 h$ \kms$\mathrm{Mpc}^{-1}$. 

\section{Data and Methods of Analysis}\label{datamocks}
Here we briefly present the SDSS LRGs
as well as mock realizations used for 
testing systematics and measurement uncertainties. 
In depth descriptions of the data and mock catalogs used here, 
as well of methods of analysis, 
are given in our previous study
of the monopole (\Kpaperalt).

In Table 1 we summarize 
the different volumes discussed in this study.


\subsection{SDSS-II LRGs}\label{data}
We use the LRG sample from the final release (DR7) 
of the SDSS.

In what follows, DR7-Full is defined as the full range of the 
SDSS LRG sample ($0.16<z<0.47$). 
We also define a  
subsample \qvlii, which 
focuses on the quasi-volume-limited region 
($z<0.36$; see Figures 1, 2 in \Kpaperalt,    
in which the latter is called there DR7-Dim\footnote{\qvl (DR7-Dim) is not a dimmer sample
of galaxies than DR7-Full, but a subsample limited 
by $z<0.36$. The term ``dim'' was used in our 
previous study to distinguish from a brighter 
overlapping subsample of DR7-Full.}). 

We calculate $\xi$ by using 
the \cite{landy93a} estimator, which 
requires the use of a catalog of random points.
For DR7-Full we use $15$ random points 
for each LRG, and for \qvl we use $50$. 

The LRG data set and random points used here are accessible on the World Wide Web.\footnote{http://cosmo.nyu.edu/$\sim$eak306/SDSS-LRG.html}

\subsection{Mock LRGs}\label{mocks}
To predict $\xi$ and its uncertainties 
in three different volumes 
(\qvlii, DR7-Full and BOSS), 
we make use of mock realizations for each volume. 
For \qvl we use mocks provided 
by the LasDamas collaboration (McBride et al., in preparation),
and for the other two samples we use mocks 
 generated by 
the Horizon Run  (\citealt{kim09a}).

The LasDamas simulations use 
a cosmology of 
[$\Omega_{M0},\Omega_{b0},n_s$,$h$,$\sigma_8$]=[0.25,0.04,1,0.7,0.8] 
and the 
Horizon Run uses [0.26,0.044,0.96,0.72,0.8], 
where 
$\Omega_{b0}$ is the present baryonic 
density and $n_s$ is the spectral index. 
Both these cosmologies are well motivated 
by constraints obtained by WMAP 5-year 
measurements of temperature 
fluctuations in the cosmic microwave 
background (\citealt{komatsu09a}).

The LasDamas collaboration 
provides very realistic LRG mock catalogs\footnote{http://lss.phy.vanderbilt.edu/lasdamas/}  
by placing galaxies inside dark matter halos using a
Halo Occupation Distribution (HOD; \citealt{berlind02a}). 
HOD parameters were chosen to reproduce the observed number density 
as well as the projected two-point correlation function $w_p(r_p)$
 of galaxies at separations $0.3<r_p<30$\hmpcii, 
 far below the scales considered here, 
 and independent of the line-of-sight clustering.  
 We use 160 LRG mock light-cone $z$-space realizations 
 that have a matching radial selection function and angular mask 
 to that of \qvlii.

For reasons explained in \S \ref{dr7dim_analysis}, 
we analyze all volumes both in 
$z-$space and real-space.
The LasDamas real-comoving-space 
catalog is similar to the redshift-space
catalog in all aspects, except the shift in $z$ due to peculiar
velocities. 
Another small difference is that 
we do not mock the observed comoving density $n(z)$ by diluting 
the original mock sample, but use the whole real-space 
catalog. In Appendix A.2 and Figure 13 of \Kpaper 
we discuss the negligible 
differences between a sample with the observed $n(z)$ 
to that originally provided by LasDamas.
For both redshift and real-space we use a ratio of $\sim 50$ random points 
for each mock data.

The Horizon Run\footnote{http://astro.kias.re.kr/Horizon-Run/} 
provides a catalog of $32$ BOSS volume realizations 
of mock LRGs with a higher 
number density than DR7, as expected in BOSS 
($n\sim 3 \cdot 10^{-4}$$h^3$ Mpc$^{-3}$ in BOSS, 
which drops quickly after $z\sim 0.6$, 
vs.~$\sim 0.9 \cdot 10^{-4}$$h^3$ Mpc$^{-3}$ in \qvlii).
LRG positions are determined by identifying
physically self-bound dark matter sub-halos that are not tidally
disrupted by larger structures.
Our only manipulations of the real-space catalogs
are to divide each of their eight full sky samples 
into four quadrants each 
to map real-space into redshift 
space, and to limit the samples to the expected volume-limited 
region of the BOSS LRGs $(0.16<z<0.6)$.  
This results in $8\times 4=32$ BOSS mock realizations.
For the BOSS volume analysis we use $\sim 2$ random points 
per mock data in $z$-space and real-space.

For our DR7-Full volume analysis   
we limit each realization to the range $0.16<z<0.47$. 
DR7-Full has 
a flux-limited region ($0.36<z<0.47$; see Figure 2 in \Kpaperalt), 
meaning that
as $z$ increases the sample 
is more biased towards the more luminous LRGs. 
This also affects the number counts of galaxies,
meaning an increase in shot noise. 

We attempt to take these two 
effects into account by subsampling the original 
Horizon Run catalog to fit the observed selection function. 
The Horizon Run team provides halo masses, 
which  we use in  
Equation 3 from \cite{park2007a} to subsample.  
In each realization we limit ourselves to $7908$ deg$^2$ 
to match the SDSS volume of DR7-Full.
The number count of the mock halos is similar to that 
of the LRGs.
For the DR7-Full volume analysis we use $\sim 10$ random points 
per mock data.

\begin{figure}[htp]
\begin{center}
\epsscale{2.5}
\plottwo{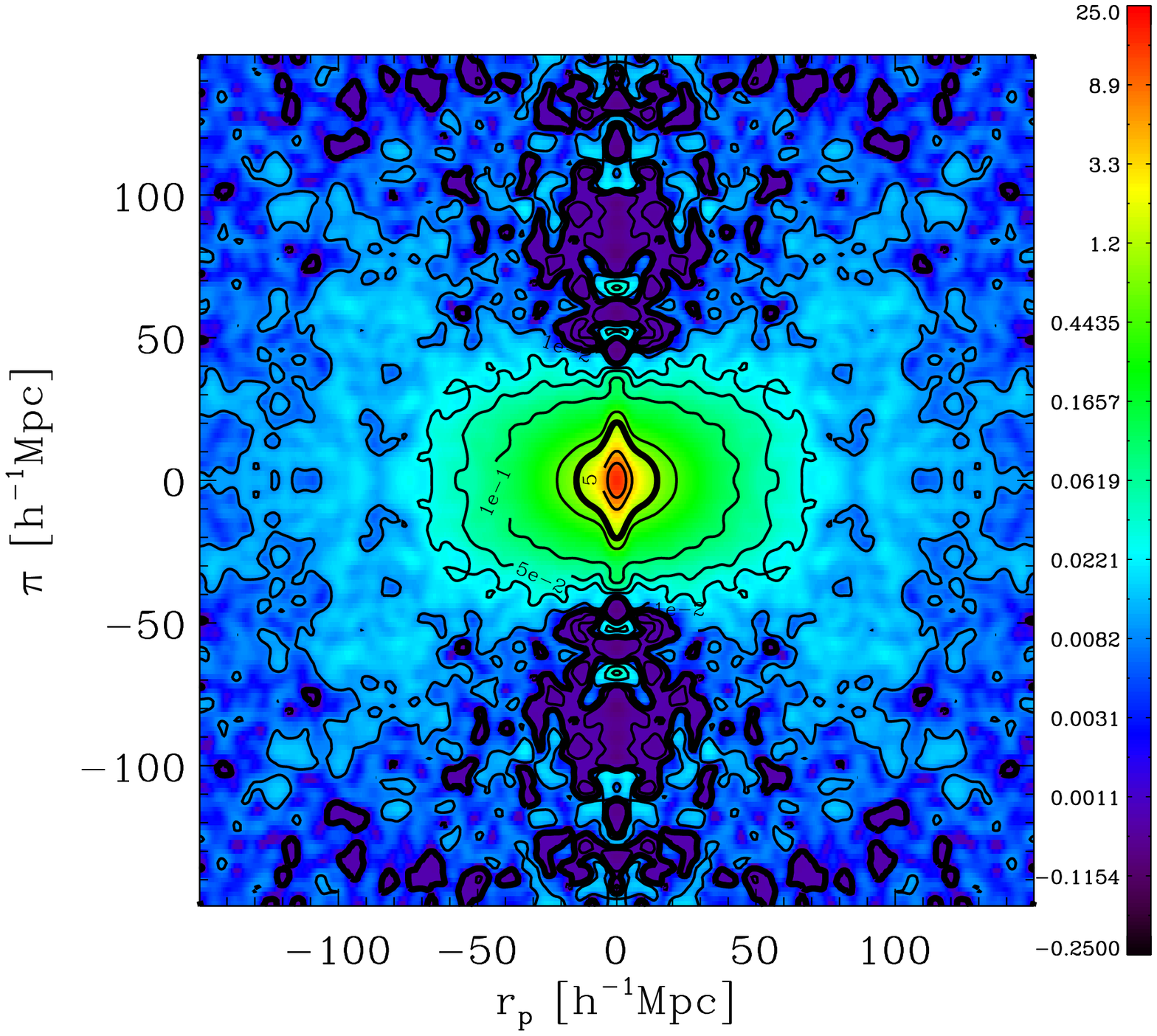}{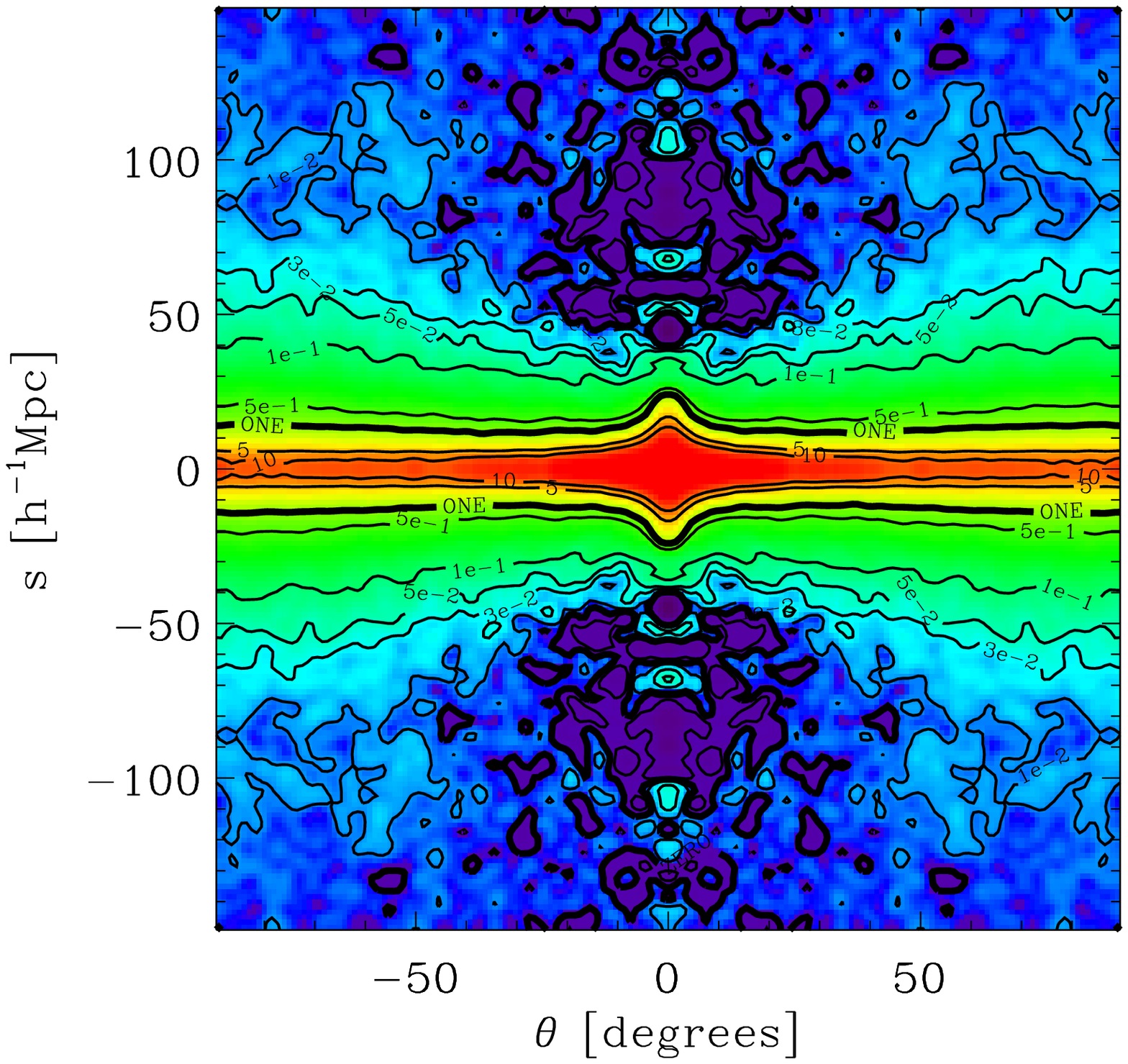}
\caption{DR7-Full Anisotropic $\xi(\vec{s})$.
{\small
Top panel shows the
$\pi-r_p$ plane; in these coordinates 
redshift distortions are deviations from circles; 
Bottom panel shows the
$s-\theta$ plane;  redshift 
distortions are deviations from horizontal lines.
$\theta=0^\circ$ is the line-of-sight direction. 
The color coding is the same for both, 
where the strongest signal is red 
and the purple 
is a negative region. 
Contour lines indicate values of
$10, \ 5, \ 1$ (thick), 
0.5, 0.1, 0.05, 0.03, 0.01, 0 (thick), and $-0.01$. 
For the purposes of these plots, we have smoothed the correlation
function using a 
Gaussian filter with $\sigma=5$\hmpc in distance, 
and with $\sigma=3^\circ$ in angle.
}
} 
\label{a2pcf}
\end{center}
\end{figure}

\section{Results: Anisotropic Clustering $\xi(\mathrm{2D})$}\label{a2pcf_analysis}

In Figure \ref{a2pcf} we show the 
anisotropic $\xi$ of DR7-Full. 
Redshift distortions due to peculiar 
velocities are apparent.
On small
scales the velocity-dispersion
effect dominates line-of-sight clustering 
(\citealt{jackson72}),  
and on large scales gravitational infall
causes a squashing effect (\citealt{kaiser87}) 
that distorts the contours 
towards smaller scales along the line-of-sight.

The top panel shows $\xi(r_p,\pi)$ in the standard 
coordinates: $\pi$ is the line-of-sight component 
of pair separation $s$;
$r_p$ is the transverse component. 
Redshift distortions in this logarithmic contour plot 
appear as deviations from circles. 
The effect of velocity dispersion is clearly seen as the 
feature at small $r_p$. On larger $\pi$ scales the
squashing effect is evident. Two notable features
are the ``negative sea'' on the line-of-sight
(also shown by \citealt{cabre09i}) and the baryonic acoustic
ridge (\citealt{okumura08}, \Gpaperii).

The bottom panel shows the same information in a 
different coordinate system. We define $s$ as the 
separation length in redshift-space: 
\beq
s=\sqrt{\pi^2+r_p^2}, 
\eeq 
and $\theta$ is the polar angle from the line-of-sight direction, i.e,
$\cos(\theta)=\pi/s$.  In these coordinates, redshift distortions
appear as deviations from horizontal lines. In Kazin et al. (in
preparation) we give a full description of our angular analysis
methods when examining the distortions on scales $s<80$\hmpcii.
Briefly, when counting pairs, we define the line-of-sight direction
($\theta=0^\circ$) as the vector that bisects  the pair 
separation vector, and
all four quadrants are binned into one. We mirror this quadrant
symmetrically for presentation purposes. For clarity, these plots have
been smoothed using a Gaussian filter with $\sigma=5$\hmpc in $r_p$,
$\pi$ and $s$ (both panels) and $\sigma=3^\circ$ in $\theta$ (bottom
panel).

We turn our focus to the region containing the
line-of-sight \baf reported by \Gpaperii.
In Figure \ref{a2pcf} this region appears as the bright positive 
spot (cyan) at 
$s (\theta\sim 0^\circ)= 100$\hmpc in the bottom panel  
(or alternatively at $\pi(r_p \sim 0$\hmpcii$)$$\sim 100$\hmpc in the top).
These plots show a sharp bright clustering excess with
$\xi>0.05$ at larger scales than that of the negative sea (purple). 

These plots can be misleading, due to the smoothing, 
so we now focus on one dimensional 
angular cuts.  

Before performing a direct comparison 
with results obtained by \Gpaper 
on the full sample ($0.16<z<0.47$;  \S \ref{dr7full_analysis}), 
which is flux limited, 
in the next section we focus on 
a quasi-volume limited sample ($0.16<z<0.36$).

\section{Results: \qvl ($0.16<z<0.36$) Line-of-Sight Clustering}\label{dr7dim_analysis}
In their Figure 15, 
\Gpaper show strong  
line-of-sight clustering measurements
where we expect 
to detect the \bafii.
Here we perform a similar procedure 
 to reproduce their results. 
 Two main differences are:  
 (1) They examine $0.15<z<0.30$, 
 where we probe $0.16<z<0.36$; 
 (2) We analyze here  both the s-$\theta$ plane, 
 and $\pi-r_\mathrm{p}$.

In this section, 
when analyzing the $s-\theta$ plane, 
we define  {\it{line-of-sight clustering}} 
as $\left<\xi(s,0^\circ<\theta<\theta_{\mathrm{max}}) \right>$, 
where
\beq\label{xisa}
\left<\xi(s,\theta_{\mathrm{range}})\right>=\int_{\theta_{\mathrm{min}}}^{\theta_{\mathrm{max}}}\xi(s,\theta)\sin(\theta)d\theta 
\eeq

In practice we calculate $\left<\xi(s,\theta_{\mathrm{range}})\right>$ 
by counting all pairs within the bin of 
dimensions $\Delta \theta$ and $\Delta s$. 

When analyzing the $\pi-r_\mathrm{p}$ plane 
we define 
line-of-sight clustering, 
as in \Gpaperii, 
as $\left<\xi(\pi,r_p^{\mathrm{min}}<r_p<r_p^{\mathrm{max}}) \right>$.
In practice this means that we count all pairs within 
these bins, and apply our $\xi$ estimator. 
In order to avoid fiber collision effects, 
\Gpaper limit themselves to  $r_p$ region $[0.5,5.5]$\hmpcii.
We test effects of fiber collisions  
by comparing data results with and without fiber collision 
corrections in weighting when counting galaxy pairs, 
as well as  with and without region $r_p<0.5$\hmpc 
and find no significant difference in results 
(see \Kpaperalt \ for details on fiber collision 
effects on the correlation function monopole)
 
Here we use  
$\theta_{max}=3^\circ$, 
corresponding to $0.5<r_p<5.5$\hmpc at $s \sim 100$\hmpcii.
This choice means that 
one expects similar line-of-sight 
clustering measurements and uncertainties at scales of 
the \bafii.  
In  \S \ref{boss} 
we discuss. similarities and 
differences between these 
coordinate systems.

\begin{figure*}[htp]
\begin{center}
\epsscale{2.55}
\plottwo{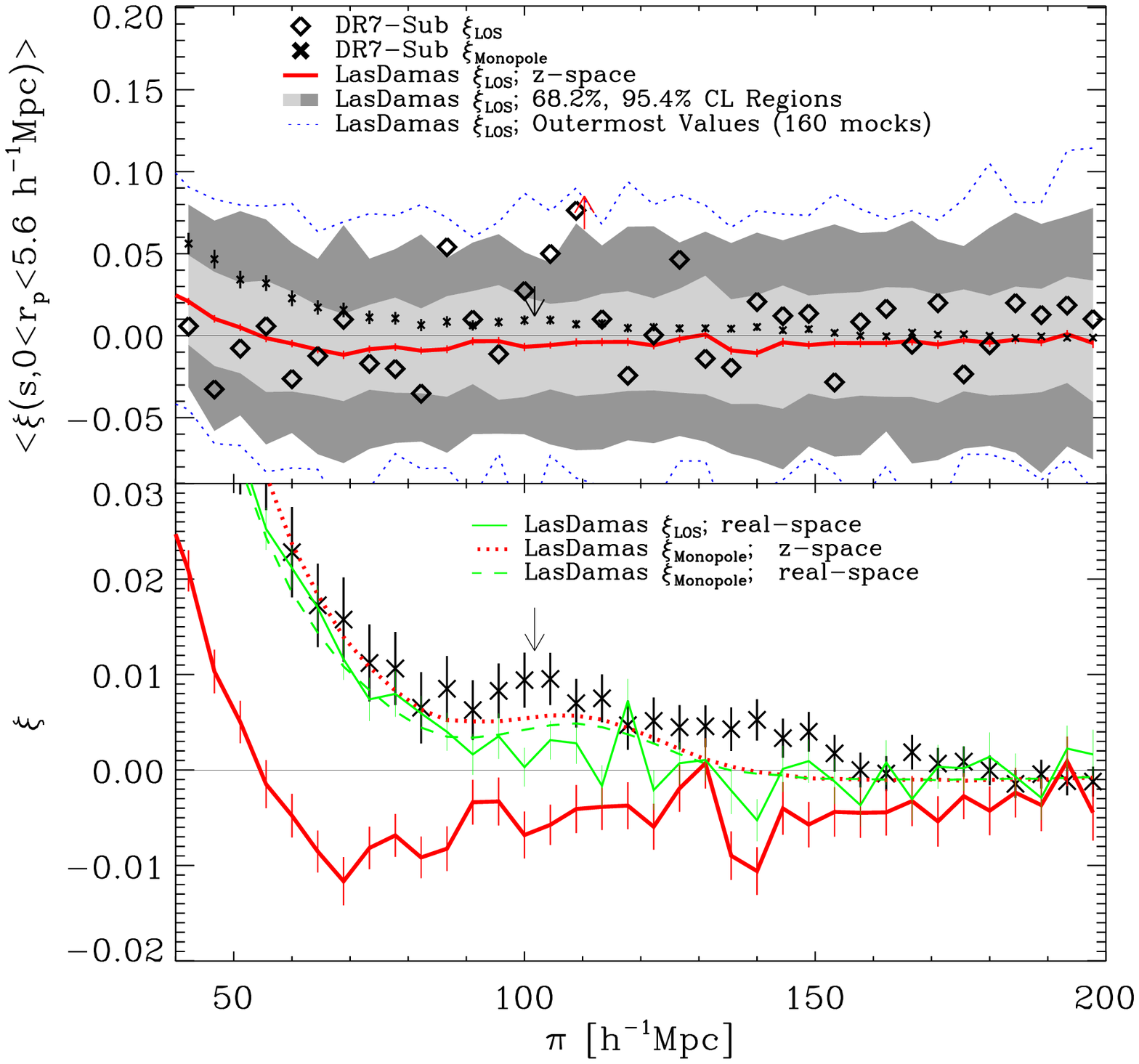}{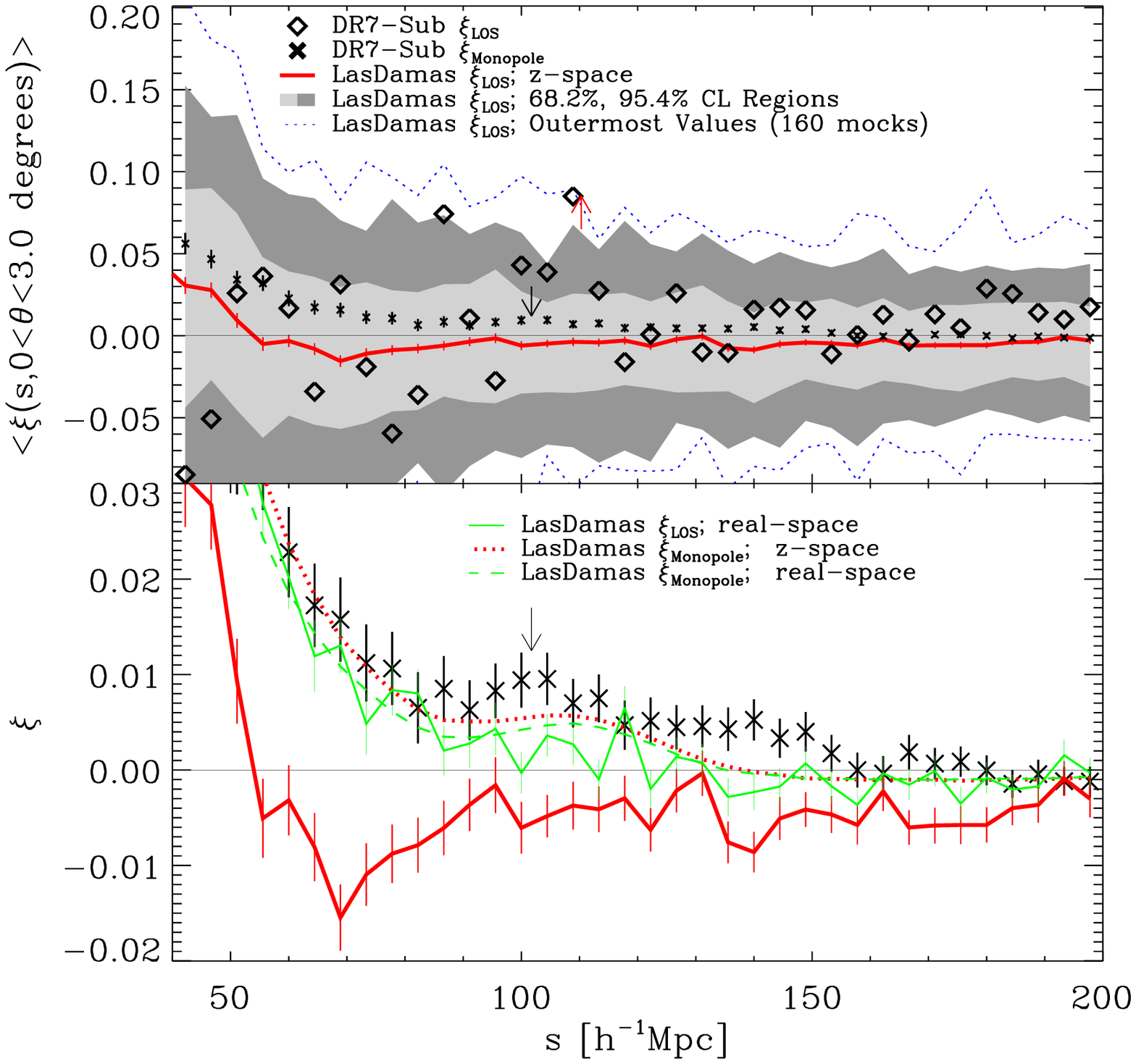}
\caption{
\qvl line-of-sight $\xi$. 
{\small
Both plots contain same information, but 
in different coordinate systems. 
On the top $\xi_{\mathrm{LOS}}(\pi,r_\mathrm{p}<5.6$\hmpcii),
and on the bottom 
$\xi_{\mathrm{LOS}}(s,\theta<3^\circ)$, as defined in the text. 
The monopoles are the same for both.
Top Panels: 
Diamonds are the SDSS \qvl LOS result. 
The crosses are the 
monopole $\xi_\mathrm{Monopole}$ for comparison. 
The solid red line is the LasDamas mock mean line-of-sight correlation
function (with 
uncertainties indicating variance of the mock mean). 
The bright and dark gray bands indicate the 
$68.2\%$ and $95.4\%$ CL regions, respectively,
for a given \qvl volume. 
The blue dotted lines are the 
outer-most values at each scale for all 160 mocks. 
The black downward arrow indicates the \baf peak location in the monopole.
The red arrow upward is where \Gpaper 
claim to detect a ``peak position'' in the line-of-sight direction. 
Bottom Panel: 
testing LOS real and $z$-space vs monopole.
The crosses  
and the thick solid lines are the same as before. 
The red dotted line
is the mock monopole in redshift-space,
and the green dashed line is the mock monopole in real-space. 
The thin solid green line is mean real-space line-of-sight signal from
all the mocks. 
The black downward arrow again shows the
monopole peak position according to \Kpaperii.
}
}
\label{LOS_SDSS_LD}
\end{center}
\end{figure*}

Figure \ref{LOS_SDSS_LD}  displays 
our line-of-sight results in both 
coordinate systems. 
In the top panels of both plots, 
\qvl line-of-sight clustering 
 results are shown by the black diamonds.  

As in Figure \ref{a2pcf}, 
the region between $50$ and $100$\hmpc 
is mostly negative. 

At scales of $\sim 110$\hmpc  
 (adjacent to the red upward arrow, which indicates 
the line-of-sight peak position according to \Gpaperii) 
we see a strong line-of-sight clustering excess in \qvlii, 
 which is much stronger than the \baf  
 in the monopole (crosses; \Kpaperalt). 
 Notice the agreement, as expected,  
between the high excess at scale $108.9$\hmpc in both choices of 
coordinate systems. 

In the top panels of Figure \ref{LOS_SDSS_LD} we 
show the expected redshift 
line-of-sight result (solid 
red line with uncertainty bars on the mean of 160 realizations), 
as well as the $1\sigma$ ($68\%$ C.L; bright gray band) 
and $2\sigma$ ($95\%$ C.L; dark gray) regions for a single 
\qvl sample. 
The blue dotted 
lines are not one realization in particular, but the outermost values  
of all mocks. 
The gray bands show 
the large scatter around the expected mean.

The expected (mock mean) line-of-sight \baf is 
not obvious,  
but appears suppressed and smeared.
This lack of clear detection might 
result from the limited statistical power available 
even from 160 mock catalogs. 

Redshift distortions  
also  
weaken the signal.
To
evaluate their  importance, we
examine the line-of-sight $\xi$ obtained
from the real-space LasDamas mocks, which are not affected by the
peculiar velocities. 

In the bottom panels of Figure \ref{LOS_SDSS_LD}, 
we compare the expected line-of-sight signal in the LasDamas 
real-space  (thin green solid line) 
and redshift-space (thick red solid line; same as top panel)
to the monopoles to which they each contribute 
(thin green dashed and thick red dotted, respectively).  
The data monopole is the same as in the top panel.

The real-space line-of-sight mock mean 
traces the monopole very well 
until $\sim 30$\hmpc (not shown here) 
and continues 
with a similar trend, though 
with considerable noise. 
Notice that at $\sim 110$\hmpc 
the $\pi-r_p$ system (top panel) 
and $s-\theta$ (bottom)  
both show signals 
appear similar to 
the monopole, but with more noise. 
There is an indication 
of a peak, but it is not obvious. 

We remind the reader that the 
uncertainty bars on the line-of-sight mock signals 
(solid green and red) 
are for the mock 
mean (of $160$), not one for one \qvl volume. 
This means that, even given a volume $100$ 
times that of the  
\qvl sample, a line-of-sight 
peak (as defined here) would 
not be obvious in real-space 
and in redshift-space it would be totally washed out. 
In \S \ref{boss2} we show that by 
using wider line-of-sight wedges 
the signal-to-noise increases, 
yielding an apparent feature.

As another consistency check, 
we perform the same line-of-sight 
test on a mock catalog with larger volume and higher density 
(Horizon Run; $z<0.6$; see \S\ref{mocks}). 
Using $32$ realizations of real-space catalogs, we 
obtain a mock mean of
$\left<\xi(s,0^\circ<\theta<3^\circ) \right>$
that mimics the monopole with 
\baf peak positions in fair agreement.  

We also estimate uncertainties with jackknife subsampling of the
data, using 24 equal area subsamples (as used in \Kpaperalt). The
jackknife uncertainties are similar to those determined 
from 
the mock
catalogs (Figure \ref{sn}). 

To test the consistency of the \qvl measurement  
with the mock-mean value we perform 
a $\chi^2$ fit, where the 
(noisy but invertible) 
covariance 
matrix is constructed from all 160 
mock realizations (data point uncertainties 
are correlated).
Examining the range $40$--$140$\hmpc  
we obtain 
$\chi^2=30.6 \ (23.2)$ for $23 \ (20)$ 
degrees of freedom (dof) 
when defining line-of-sight as $r_p<5.5$\hmpcii,
and 
$\chi^2=31.7 \ (24.8)$ 
when using $\theta<3^\circ$.  
A $\xi=0$ model yields 
$\chi^2=26.5 \ (23.2)$ and $27.0 \ (29.0)$. 
For these measurements 
we use binning 
$\Delta \pi=4.4 \ (5.0)$\hmpc and 
$\Delta s=4.4 \ (5.0)$\hmpcii,  
respectively. 

These results show that  
when using the smaller
binning of $4.4$\hmpcii, 
all these tests indicate a $\sim 1\sigma-1.5 \sigma$ 
agreement between the data and the expected result, 
and even agree even better with a null hypothesis. 
Widening the bins  ($5.0$\hmpcii) yields 
even better agreements.


When restricting the test around the expected \baf 
at $100-125$\hmpc we obtain 
$\chi^2=11.2 \ (13.6)$ for $6 \ (5)$ dof, 
and 
$\chi^2=11.3 \ (13.4) $ for $r_p$ and $\theta$ 
line-of-sight definition, respectively.
 A $\xi=0$ model yields 
$\chi^2=9.9 \ (12.1)$ and $10.1 \ (11.75)$.

When widening the bins  ($5.0$\hmpcii), the 
null hypothesis yields better results than the 
observation, 
though the data agrees only at a $\sim 2 \sigma$ 
level.  


These tests show that the 
line-of-sight observation is 
in 
good
agreement with 
a  fiducial $\Lambda$CDM model.  

Next section 
we perform the same analysis on DR7-Full,
which has a much larger volume. 
We also compare results directly with \Gpaperii.

\section{Results: DR7-Full ($0.16<z<0.47$) Line-of-Sight Clustering}\label{dr7full_analysis}

\begin{figure*}[htp]
\begin{center}
\epsscale{2.}
\plotone{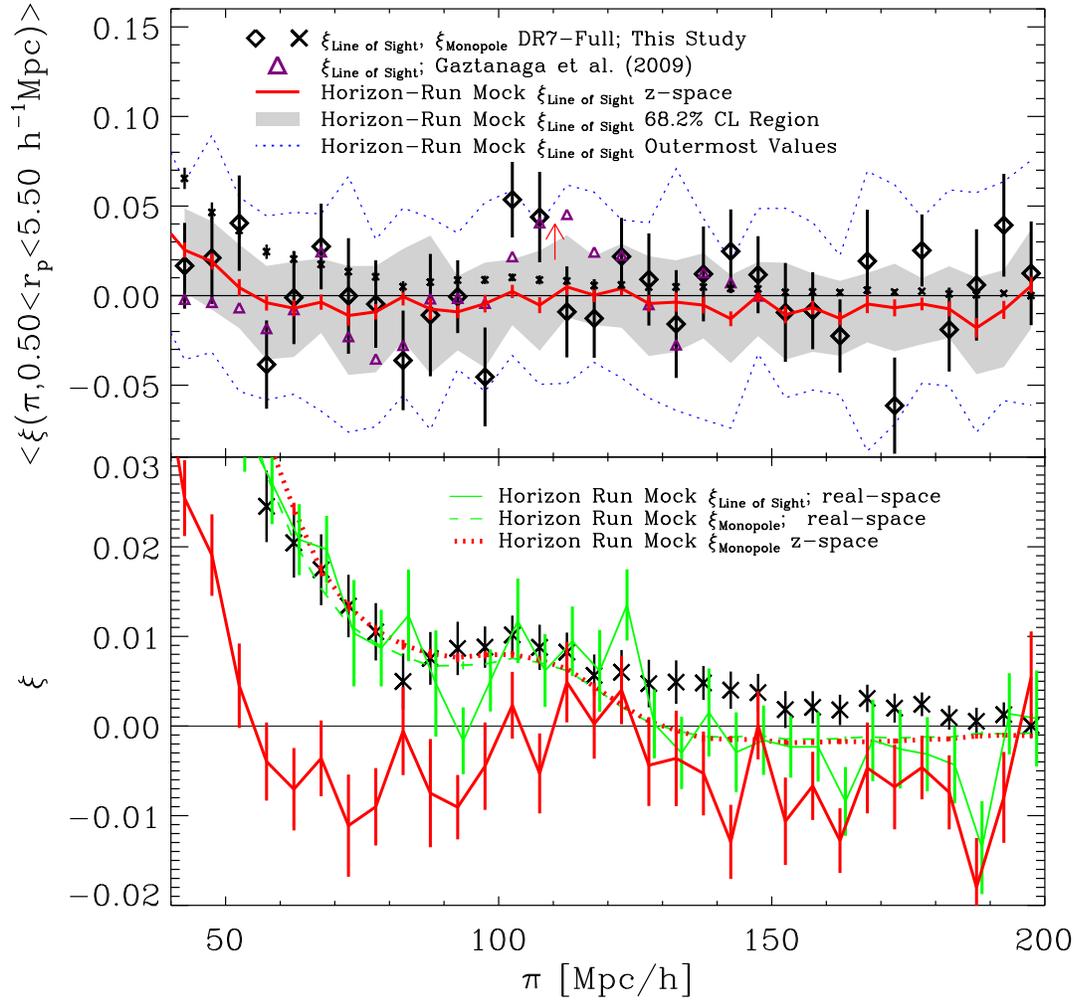}

\caption{
DR7-Full line-of-sight $\xi(\pi,r_p<5.6$\hmpcii).   
{\small
The legend here is the same as in Figure \ref{LOS_SDSS_LD} 
with a few differences- 
mock results are based on $32$ Horizon Run mocks, 
we add results from \Gpaper (purple triangles) 
for direct comparison
}
}

\label{los_dr7full_plot}
\end{center}
\end{figure*}

In the top panel of Figure \ref{los_dr7full_plot} we show a direct 
comparison between our DR7-Full 
line-of-sight clustering $\xi_{\mathrm{LOS}}$ 
results (black diamonds)  
to those in Table 1 of  \Gpaper 
and their Figure 15 
(purple triangles; DR6). 
The $1\sigma$ uncertainties 
on our data are explained below.
We omit uncertainties of \Gpaper 
 to avoid cluttering, but show 
 in Figure \ref{sn}  
 that our uncertainties 
 are similar to theirs  
 on these scales.

As expected, the measurement is very noisy.  
Our results are similar to those obtained by \Gpaperii, in two
ways. 
First, we both measure the negative 
region between $50-95$\hmpcii, 
and second, we each measure a 
{\it
very 
}
 high value
that happens to be where one 
expects the acoustic feature to be. 
This strong excess at $\pi \sim 110 $\hmpc 
appears five times stronger 
than that obtained in the much 
higher signal-to-noise ratio monopole (crosses; \Kpaperalt). 

In \S \ref{mocks} we explain the technical details of diluting 
the original Horizon Run mocks (which were originally 
designed for BOSS volume and density), to
the properties of DR7-Full. 
The resulting expected $z$-space monopole is shown as 
the red dotted line in the bottom panel of Figure  \ref{los_dr7full_plot}.
The agreement with the data around the 
\baf seems good. However, we do notice a $\sim 10\%$ bias difference 
 $(\sqrt{\xi_\mathrm{mocks}/\xi_\mathrm{SDSS}})$ 
 at smaller scales. 
 This might indicate that we have excluded a few more
 low mass halos than we should have, 
 since clustering is known correlate with mass 
 (\citealt{zehavi05a}). 
 We doubt 
 that this mismatch will affect our line-of-sight analysis around 
 the \bafii. 
 In the previous section we 
 obtained similar conclusions 
 to those presented here when using 
 very realistic mocks provided 
 by LasDamas, that are fit to data on low scales.
In the bottom panel of Figure \ref{sn} 
we also show excellent agreement between 
our uncertainty measurements
and those obtained by \Gpaperii.

The predicted line-of-sight correlation function in redshift-space
is shown in both panels of Figure \ref{los_dr7full_plot} 
as the solid thick red lines. 
The bottom panel is a zoomed-in version 
of the top. The uncertainty bars indicate 
the uncertainty in the mock mean given $32$ realizations.  
In the top panel of Figure \ref{los_dr7full_plot}, 
the gray band indicates 
the $68.2\%$ confidence level (CL) region for a 
single volume, and the blue dotted lines are the outermost 
values of all mocks (not one in particular). 

We see that the 
expected line-of-sight correlation function 
has a negative valley around $55$--$100$\hmpcii. 
The signal increases towards the \baf area and decreases on larger
scales.  

Comparing these predictions to observation, 
we see that the very strong measurement at $\pi\sim 100$\hmpcii, 
although unlikely, is still acceptable 
according to the mock realizations.

We note that, although our measurements are very similar,  
we do not obtain the exact same results as \Gpaperii.  
Differences might result 
from any combination of systematics in analysis. 
For example, our
methods of  
weighting galaxies may slightly differ when counting pairs, 
or perhaps it may be due to the differences 
in the samples (we use $\sim 105,000$ LRGs where they use $\sim 77,000$). 
When comparing the much more stable monopole we obtain similar results
to theirs. As for the anisotropic $\xi$ 
we find that we obtain a similar normalized 
quadrupole $Q(s)$ to that published in \cite{cabre09i}. 

Our line-of-sight $\xi$ results, as well as covariance matrices, 
may be obtained on the World Wide Web.\footnote{http://cosmo.nyu.edu/$\sim$eak306/BAF.html}

\section{Interpretation of Data Results}\label{interpretation}
In the previous section we show 
that our measurements are very similar 
to those previously obtained by  \Gpaper and in 
see Figure \ref{sn} we show agreement in uncertainty estimates. 
These agreements  
show that the  
results are not sensitive to 
minor systematic differences (selection of LRGs, weighting algorithms, etc.). 
We do, however, disagree on the interpretation 
of the results, as we now clarify. 
The main distinction between the 
two interpretations is that 
we do not agree on the importance of 
a null $\xi=0$ test. They claim 
that it is 
{\it ``only slightly disfavored compared to the best-fit model"}. 
Here we compare the null test 
to physical models and 
conclude that the $\xi=0$ is a 
good fit to the data, and physical models 
do not perform significantly better. 
This means that 
the line-of-sight data 
{\it alone} 
is too noisy to 
infer the presence of a peak. 

\Gpaper argue for a significant detection 
of the line-of-sight \baf 
based on 
$\chi^2$ fits to the data (see their Figures 13 and 15).  
Here we investigate the significance of 
their results compared to ours. 

In Table \ref{jeffreys} we summarize the $\chi^2$ results 
they and we obtain with various models on 
similar data sets. We analyze \qvl and DR7-Full,
 while \Gpaper investigate the full DR6 as well 
 as a smaller volume ($0.15<z<0.30$).  
In all cases the  line-of-sight $\xi$($\pi, r_p<5.5$\hmpcii) 
is investigated at scales of $40<\pi<140$\hmpc  
with bin widths of $\Delta\pi=5$\hmpc  ($N=20$ bins). 

The models that \Gpaper investigate are:
\begin{enumerate}
\item BAO: best fit $\Lambda$CDM based line-of-sight model based on $k=5$ parameters. 
\item BAO+mag: same as BAO including lensing magnification effect ($k=6$).
\item No BAO: a $\Lambda$CDM model based on a featureless P ($k=5$).
\item $\xi=0$: Null test ($k=0$). 
\end{enumerate}

We investigate two fiducial flat $\Lambda$CDM models based on our mocks (LasDamas, Horizon Run). 
These models are very similar ($\Omega_{M0}=0.25,0.26$, respectively). 
As we do not vary the cosmology in our estimates, here we use $k=0$. 
Both use $N$-body simulations 
that produce very realistic mock LRG catalogs. 
As described in \S\ref{mocks},  
the mocks take many observational effects into account. 
Assuming the correctness of $\Lambda$CDM, 
this procedure yields very reliable uncertainties ($C_{ij}$), 
as well as non-linear fiducial models.  
The models used by \Gpaper are analytical  
(giving them the advantage of probing many 
cosmological models), but they  
do not account for effects of the survey mask  
in their modeling or in their $C_{ij}$. 

We also investigate the $\xi=0$ model,
 which is, of course,
not a physical one. 
It is, however, an interesting straw-man 
model in the context of claiming a detection.

We test our models on both our data, 
and those in Table 1 of \Gpaper (with our own covariance matrices). 
 
 Examining the reduced $\chi^2$  column
 ($\chi^2/(N-k)$) in Table \ref{jeffreys} , we 
 conclude that the data
 agrees fairly well with all 
 the models tested. 
 We see this in the $Prob$ 
 column which indicates 
 the probability of a random variable 
 from a $\chi^2$ distribution 
 with $N-k$ degrees of freedom 
 to 
 have a value larger than that 
 in the $\chi^2$ column. 
 Notice that all models 
 range betwen $4.5\%-96\%$. 
 In other words, all models 
 fit the data at $2\sigma$ or better. 
 This procedure tests each model independently, 
but to compare between them is more complicated. 
  
There are various ways to 
compare models and  
determine significance of 
difference. Here we 
present a few common tests 
used in the literature (\citealt{liddle09a} and references within). 

The Jeffreys scale  uses a value called  
the ``Evidence" $E$, 
which 
is the average likelihood 
of the parameters averaged over the 
parameter prior. 
The difference of $\ln E$ may 
be used to describe how 
much better one model agrees 
with data from another.
For example, in order for one 
model to perform significantly better 
than another, $\Delta \ln E$ should 
be larger than unity. 
Other useful 
divisions in significance is for 
values $2.5$ (posterior odds of $12:1$) and $5$  ($148:1$; see caption of Table \ref{jeffreys}). 
 
We approximate $E$  
by using the
Bayesian information criterion 
(BIC; \citealt{schwarz78a})
defined as 
\beq
-\ln(E) \sim \mathrm{BIC}=-2 \ln(L_{max})+k \ln(N) ,
\eeq
where $k$ is the number 
of parameters, $N$ is the number 
of data points. 
Assuming the liklihood $L$ 
is Gaussian $-2ln(L_{max})= \chi^2_{min}$. 
Using this criterion, we prefer models 
that yield a lower BIC result. 
A similar technique is  
the Akaike Information Criterion (AIC; \citealt{akaike74a}) 
defined by replacing  
$k \ln N$ with $2k$.  
Both these criteria are summarized 
in Table \ref{jeffreys}  in separate columns. 

Applying BIC and AIC to 
the \Gpaper results, 
the ``No BAO" model is clearly 
disfavored in respect with  
the other models. 
This result does not mean, however, 
that the data reveal a significant 
line-of-sight \bafii.  
One way to look at this is to realize that 
$\xi=0$ is a good fit. Below we elaborate on 
this point. 

For completeness we mention 
that the $\chi^2$ for the physical models 
posted in \Gpaper  might be misestimated 
for the purpose of comparing them to each other.  
When fitting models to data they use a 
Gaussian prior likelihood indicated in 
their Section $3.8.2$.  
This prior should be taken into account 
when calculating the full $\chi^2$. 
This means that an additional term 
$\chi^2_\mathrm{prior}=\sum_{i=0}^{N_\mathrm{prior}}(\gamma_i-\gamma^{\mathrm{prior}}_i)^2/\sigma^2_{\gamma_i}$, 
where $\gamma_i$ is the $i^\mathrm{th}$ parameter, 
should be added. 
They use $N_\mathrm{prior}=4$  parameters as priors: [$\Omega_{b}$, $\Omega_{M0}$, $b\sigma_8$, $\beta$], 
where $\beta$ is the Kaiser squashing parameter, 
and $b$ is the linear bias factor relating matter and LRG over-densities.  
Through private communication with 
the authors we learn that their best fit values for the BAO+mag 
model are $[0.049, \ 0.240,  1.73, 0.39]$
This yields  $\chi^2_\mathrm{prior}=9.4$ which 
should be added to their stated $\chi^2=8$, 
and correcting the number of data points from $N=20$ to $24$ (one additional ``data point" for each prior). 
They claim that they obtain similar parameter values ($\gamma_i$) for 
the other physical models, so when comparing between them  
this term should approximately cancel out.   
It is not clear, however, the correct manner to incorporate 
this correction utilizing BIC and AIC 
 when comparing with $\xi=0$ and 
our mock models which do not use  
prior likelihoods. For simplicity we quote in Table \ref{jeffreys} 
values published by \Gpaperii. 

So far we have used a rigid definition of number of parameters 
$k$. A thorough analysis would investigate 
the relative influence of each parameter 
on the model. This could be done by investigating 
the full parameter space, which is out of the 
scope of this study.  
To attempt to minimize effect of free parameters, 
we asked the authors  of \Gpaperii,  
through private communication, 
for results using the minimum number  
of free parameters.  
Setting all parameters to the prior means and 
the shift in radial scale $D_r$ to their fiducial model  they obtain 
for the BAO model $\chi^2=12.6$ ($k=0$) as listed in Table \ref{jeffreys}, 
and for BAO$+$mag they obtain $\chi^2=9.7$  ($k=1$). 
Comparing BAO to 
their $\xi=0$ ($\chi^2=14$; $k=0$) we obtain $\Delta\chi^2=1.4$, 
showing no significant improvement (slightly above $1 \sigma$). 
Comparing BAO$+$mag to $\xi=0$ we obtain BIC=$1.4$ and AIC=$2.4$. 

We also examine our Horizon Run model and $\xi=0$ to their data. 
We use our DR7-Full  covariance matrix,   
though normalize it by using the ratio of volumes of DR7 and DR6 (see caption in Table \ref{jeffreys}),  
and obtain  $\Delta\chi^2=1.4$.  
We conclude that using the DR6 data the physical models 
do not out-perform $\xi=0$ in a significant manner.

In the case of DR7-Full data 
we see that the fiducial Horizon Run 
model performs better than $\xi=0$ ($\Delta\chi^2=3.1$). 
This does not yield much confidence 
that we have detected a line-of-sight peak for a few reasons.  
First, the model \baf 
is much weaker than 
the strong clustering in the data (Figure \ref{los_dr7full_plot}). 
Second, the physical model is preferred by less than $2\sigma$. 
Third,  
The number of bins might be of concern. 
We conduct  
tests with  
wider bins ($\Delta\pi\sim 10$\hmpcii), 
and obtain 
a better agreement between the models $\Delta\chi^2=1.1$. 
This shows that the criterion is sensitive to 
the width of the binning when $k=0$.
 
We note that the $\xi=0$ 
fit to DR7-Full is much worse than that to DR6 (\Gpaper data). 
We mentioned the normalization of the $C_{ij}$;  
without this normalization we obtain only slightly better fits. 
Another notable difference that might contribute to this discrepancy 
is that  
that the \Gpaper data is much smoother than ours.  
They explain binning techniques applied on the data, 
which we do not perform here.  
Another noticeable difference are the two 
bins centered around $112.5$\hmpc and $117.5$\hmpcii. 
In \Gpaper DR6 results they are positive and form 
a wide range of positive over-density. 
In DR7 we show in Figure \ref{los_dr7full_plot}, 
as well as Figure 12 in \Gpaper (see their red lines) 
that these two bins are negative.

We emphasize here that many assumptions  
are made when performing this comparison.  
We assume $C_{ij}$ is model independent, 
but do not expect it to change significantly 
within parameter space. 
In addition, 
we assume the likelihood 
to be Gaussian and the correctness 
of the BIC and AIC. 
As noted above,   
we find sensitivity in $\Delta\chi^2$ model comparisons 
 when varying 
size of $\pi$ bins. 
 
To summarize, both studies demonstrate that 
the line-of-sight measurement  is very noisy
(in a $\chi^2$ sense). 
It is our opinion 
that these correlation functions 
do not convincingly 
show a line-of-sight feature.

The main distinctions between our conclusions 
and those of \Gpaperii, is that when investigating 
significance we take into account the 
addition of free parameters $k$, 
as well as the importance of the $\xi=0$ test. 
$k$ serves in both BIC and AIC as 
a ``penalty" for adding parameters. 
Ignoring $k$ (which is the wrong thing to do) 
would cause this test to favor a 
BAO model with magnification, 
where including it, 
this comparison shows 
that the current data can not distinguish 
between physical models in a convincing manner. 
The physical models do perform better than 
a No BAO model, 
This last point, however, is not proof for detection of a peak 
in the line-of-sight clustering. 
No BAO is not 
physically motivated, and there is 
no {\it physical} reason to prefer this model 
over other featureless models. 
We show here that the physical models 
do not perform better significantly than 
a null $\xi=0$, which yields a good fit to the data.

The reader should keep in mind 
that there is no apparent correct  
answer to the issue at hand 
for the current data. 
In order to detect 
a significant line-of-sight 
\bafii, which can be 
used to determine $H(z)$ 
and ultimately the expansion 
rate of the nearby universe, 
noise must be 
reduced by probing larger volumes.

In the following section, we examine the line-of-sight signal expected
in a much larger volume survey.

\section{Prediction: BOSS Line-of-Sight Clustering}\label{boss}

The Baryonic Oscillation Spectroscopic Survey (BOSS; \citealt{schlegel09a}) 
plans to map $1.5$ million LRGs in a 
much larger volume than DR7,
up to $z\sim 0.7$.
We examine here what can 
be expected from
the line-of-sight correlation function in the BOSS volume.

\begin{figure*}[htp]
\epsscale{2.}
\plotone{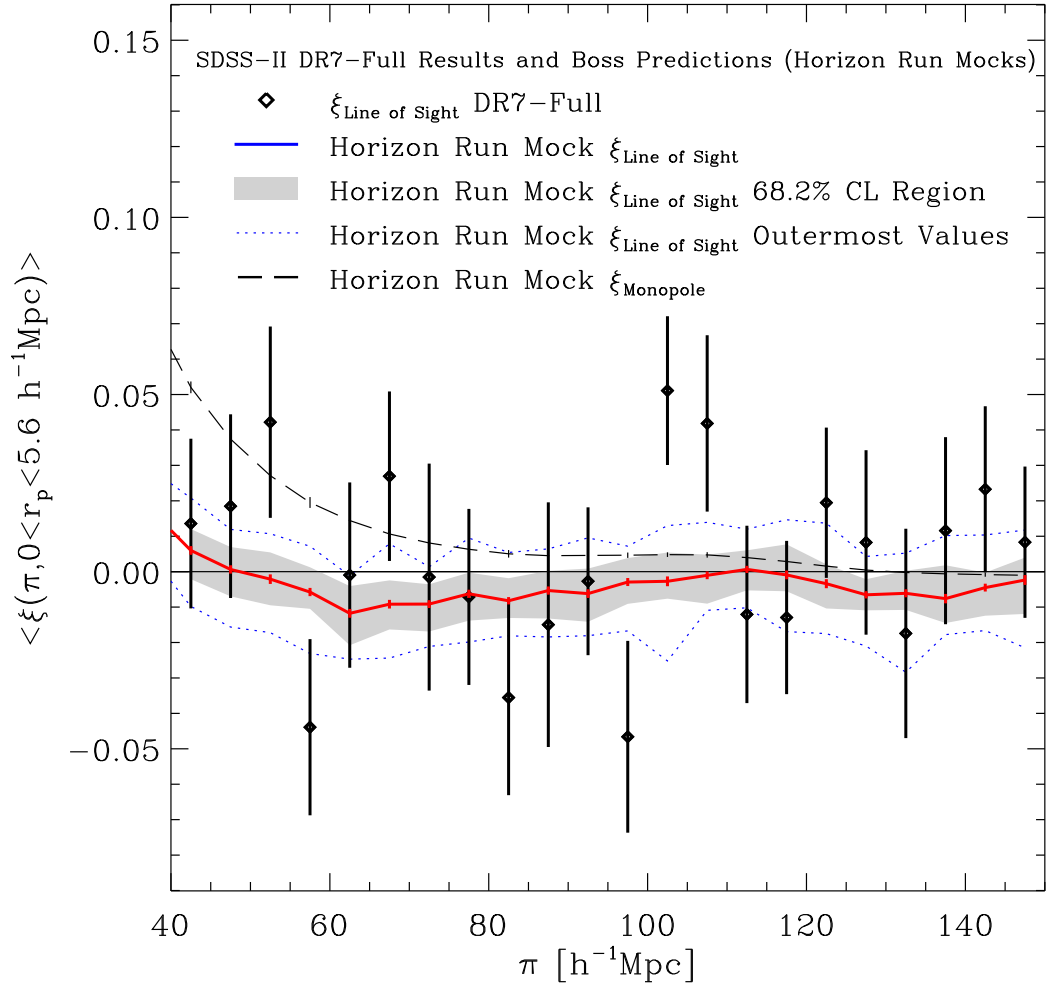}
\caption{
DR7-Full vs expected BOSS Line-of-sight $\xi(\pi,r_p<5.5$\hmpcii). 
{\small
The red line is the mean of the Horizon Run predictions for the
BOSS volume.
The 
$68.2\%$ CL region is the gray band. 
The diamonds are the SDSS-II measurement from DR7-Full 
(same as in Figure \ref{los_dr7full_plot})
The apparent high clustering excess in the current data at $\pi \sim 110$\hmpc 
should be excluded (or confirmed)  
at a high confidence level by the BOSS data. 
For comparison, the dashed line shows the mock monopole 
(with $1\sigma$ uncertainties for a given 
BOSS volume).
}
}
\label{horizon_run}
\end{figure*}

Figure \ref{horizon_run} displays the predictions for the
line-of-sight signal. In this case 
we work in $\pi$-$r_p$ space and define 
the line-of-sight 
using a cut at $r_p = 5.5$ \hmpc (using $\Delta \pi=5$\hmpc bins).
The solid red line is the line-of-sight prediction for the 
mean BOSS signal, and 
the gray band is the $1\sigma$ sample variance. 
The black thick diamonds are the observed results from DR7-Full using
the same line-of-sight definition (same values as in Figure \ref{los_dr7full_plot}).
The dashed line shows the expected monopole prediction for comparison.

Although the line-of-sight mock signal is negative 
at these scales due to the linear
redshift-space distortions (the squashing effect),
there is a signature of a
peak with a position 
in fair agreement 
with the monopole. 
The $1\sigma$ uncertainties suggest, however, 
that even in BOSS we do not expect a significant 
line-of-sight detection of the
\bafii, if we define the line-of-sight as narrowly as \Gpaper do
(that is, within $\theta< 3^\circ$ or $r_p<5.5$\hmpcii). 

The estimated uncertainties do indicate 
that BOSS will have the statistical power to 
rule out (or confirm) at high confidence a clustering excess 
at the level claimed by \Gpaper 
at $\pi \sim 110$\hmpc.
To test this proposition quantitatively, 
we evaluate $\chi^2$ 
of the 
DR7-Full result versus the
model constructed from  
the Horizon Run 
mock mean, using 
all $32$ realizations 
to determine the covariance matrix. 
We obtain 
$\chi^2=151$ for $4$ degrees of freedom 
(the bins between $100<\pi<120$\hmpcii). 
Our mocks thus predict that
a measurement as strong
as that 
seen in DR7-Full is
extremely 
unlikely in BOSS. 
 
\begin{figure*}[htp]
\epsscale{2.4}
\plotone{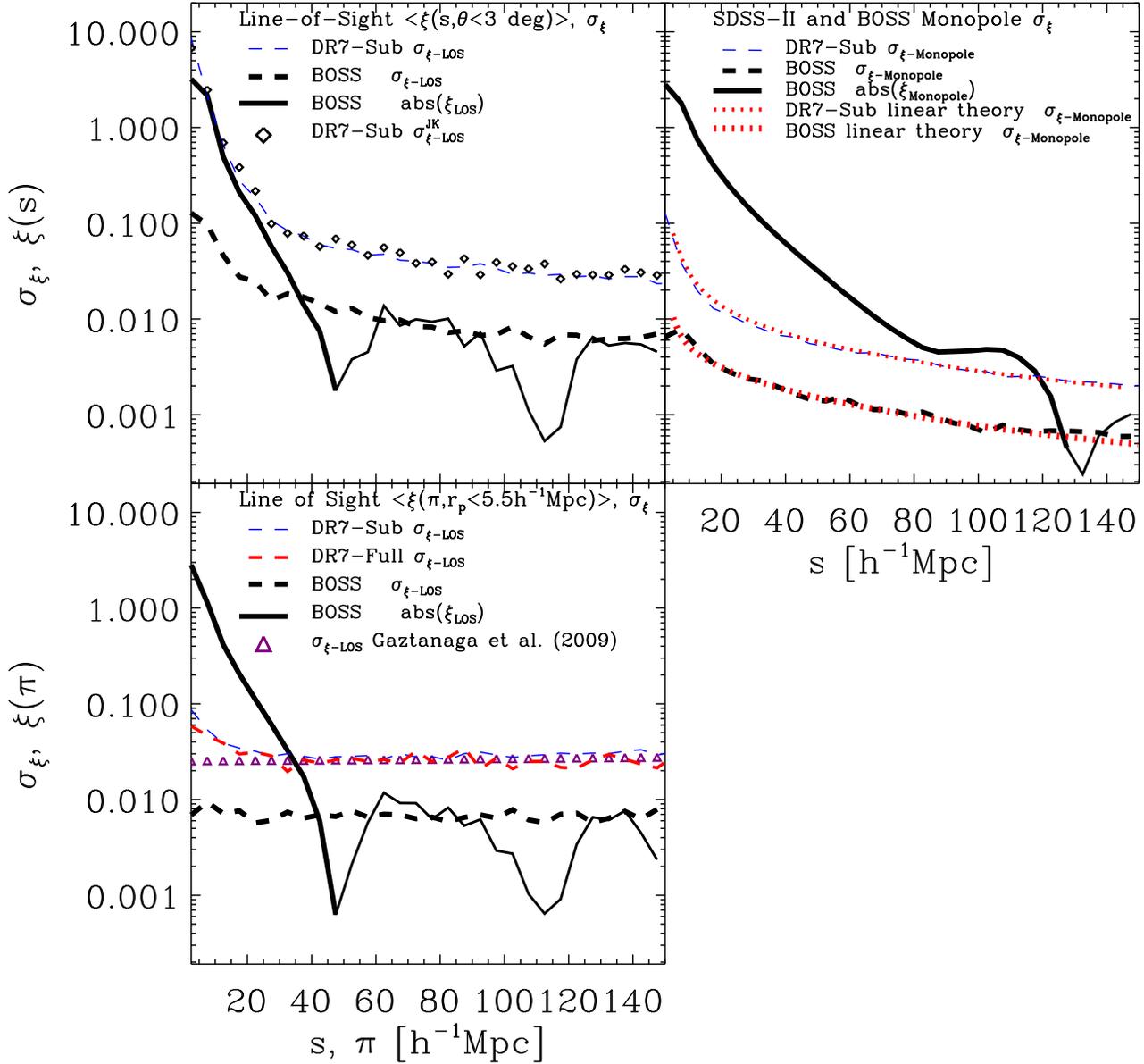}
\caption{
\qvlii, DR7-Full and BOSS line-of-sight and monopole $\xi$ and uncertainties $\sigma_\xi$. 
{\small
Top left panel shows the line-of-sight (defined as $\Delta\theta<3^\circ$). 
The solid line is $|\xi_{\mathrm{LOS}}|$ 
(where thick is positive and thin negative values). 
The dashed lines are the line-of-sight  
uncertainties  $\sigma_{\xi-\mathrm{LOS}}$ based on mock
catalogs.  The thin dashed line is for \qvl (using LasDamas mocks),
and the thick dashed line is for BOSS (using Horizon Run mocks).  
The diamonds show jackknife
uncertainty estimates for \qvl data. 
The bottom left panel is similar where 
line-of-sight is defined as $r_p<5.5$\hmpcii. 
Differences- red dashed is DR7-Full results (using Horizon Run mocks), 
which agree very well with results from \Gpaper (purple triangles). 
The right panel shows the same quantities as top left for the monopole.
The solid line is $|\xi_{\mathrm{Monopole}}|$. 
The dashed
lines are its uncertainties
$\sigma_{\xi-\mathrm{Monopole}}$ based on mock catalogs. The thin
dashed line is for \qvlii, and the thick dashed line is for
BOSS. The dotted lines show the estimates from linear theory based on
\citet{cohn06}. The thin dotted line is for \qvlii, and the thick
dotted line is for BOSS. 
}
}
\label{sn}
\end{figure*}

In Figure \ref{sn} we examine the uncertainties of the
line-of-sight (left panels)
and monopole (top right panel) 
measurements of the
correlation function.  
We examine differences between 
the two line-of-sight definitions: 
the top left panel shows for $\theta<3^\circ$, 
and the bottom  $r_p<5.5$\hmpcii.
Included in this figure are the estimates of 
the \qvl sample (both using
LasDamas and using data jackknife subsamples), 
DR6 (as estimated by \Gpaperii), 
DR7-Full (Horizon Run)
and for BOSS 
(Horizon Run). 
Finally, we also compare our monopole predictions to
those found using the linear theory estimate of \citet{cohn06}.
For descriptions of the mocks realizations used 
please refer to \S\ref{mocks}.

The top left panel of Figure \ref{sn} shows results for the
line-of-sight ($\theta<3^\circ$). 
For comparison, the solid line is the expected 
signal (in absolute value) using the BOSS mocks from the Horizon Run. 
The thick dashed line
shows the uncertainty estimates for this case, demonstrating again that even in
BOSS the line-of-sight signal, as defined here, will be very
difficult to measure.  
The thin dashed line shows our error estimates
using the LasDamas \qvl mocks. For comparison the diamonds show jackknife
estimates of the uncertainties from the data itself, which are in excellent
agreement with the mocks. Obviously, the SDSS-II results are much
noisier than the BOSS results will be.

The bottom panel shows the same predictions 
but when defining the line-of-sight region as $r_p<5.5$\hmpcii. 
The medium width dashed red line corresponds to our 
DR7-Full uncertainties, the purple triangles 
are uncertainties according to \Gpaper (see their Table 1). 
The \qvl uncertainty (thin blue dashed line),  
the predicted BOSS signal (thick solid black line) 
and uncertainty (thick dashed black line) 
has the same notation as before. 
\cite{cabre09i} argue that at $\pi>20$\hmpc 
shot noise dominates the noise. 

Comparing the BOSS results, 
we see clear differences between the uncertainties 
in the two coordinate systems. 
The $s-\theta$ has a negative slope in respect 
to scale where the $\pi-r_p$ has a very slight positive 
slope (also noticeable in Figure \ref{los_dr7full_plot}). 
The reason for this is simple: 
$\theta<3^\circ$ corresponds to a cone, 
whereas $r_p<5.5$\hmpc corresponds to a cylinder. 
The uncertainties in the former decrease with $s$ 
because more data is included, reducing the noise. 
The latter is very flat, because all scales
each bin contains roughly the same number 
of pairs. The slight positive slope 
indicates minor effects of boundary conditions at these 
scales. 
We thank E. Gazta\~{n}aga for pointing this difference out 
in a private communication. 

As a consistency check, we verified that, 
although the scale dependence differs,
the uncertainty for BOSS at 
$\sim 100$\hmpc is similar in the two coordinate systems, 
as expected. 

Our DR7-Full uncertainties 
are in excellent agreement with 
those of \Gpaperii.
This result is slightly surprising, because
we expect our DR7-Full uncertainties to 
be slightly smaller than those obtained by 
\Gpaperii, who investigate the slightly smaller DR6 volume.
This might be an artifact of the way 
we subsampled the Horizon Run mocks, as 
explained in \S \ref{mocks}.
Overall, however, we consider our result to 
be in general agreement with those of \Gpaperii.

The right panel shows results for the monopole. For comparison, the
solid line is the absolute value of the expected signal from 
the BOSS mocks. The
thin dashed line shows the uncertainty estimates from LasDamas for
SDSS-II (\qvlii).  The thick dashed line shows the uncertainty
estimates from Horizon Run for BOSS, which are obviously much
smaller. The thin and thick dotted lines show the estimates for each
survey from linear theory (\citealt{cohn06}),
which are in remarkable agreement with the mock catalog results. This
comparison demonstrates that our estimates from the mock catalogs are
reasonable and strengthens our confidence in our estimates for the
line-of-sight error estimates.

BOSS
will have a larger volume 
than \qvlii, and hence yields
much smaller uncertainties in 
$\sigma_{\xi-\mathrm{LOS}}$ (line-of-sight) 
and 
$\sigma_{\xi-\mathrm{Monopole}}$ (monopole).  
We estimate that the monopole uncertainty 
at the \baf 
will 
be reduced by a factor of four 
and the line-of-sight signal by 
approximately the same amount
We predict the signal-to-noise for BOSS at the \baf scale at
$S/N\equiv |\xi_{\mathrm{Monopole}}|/\sigma \sim 6$. 
These calculations are based on a single bin, 
so should be considered a lower limit 
of the true statistical power of the survey.  

Notice that the BOSS 
 $\sigma_{\xi-\mathrm{LOS}}$ 
(left panel of Figure \ref{sn}) is expected to 
be slightly larger than that 
of the SDSS-II  
$\sigma_{\xi-\mathrm{Monopole}}$ 
(right panel of Figure \ref{sn}). 
Moreover, the BOSS $\xi_{\mathrm{LOS}}$, 
appears negative at 
scales larger than $s>50$\hmpcii, 
resulting in a baryonic acoustic dip 
in the left panel, 
which is significantly 
smaller in absolute 
magnitude than $\sigma_{\xi-\mathrm{LOS}}$, 
making a detection of the feature unlikely. 

We emphasize that the low signal-to-noise we 
predict is a result of the very small angular 
range that \Gpaper chose to study. 
Our results should not discourage us from 
disentangling $D_{\mathrm{A}}$ and $H$ 
using the BOSS sample, as we describe in the next section. 

\section{Separating Line-of-Sight and Transverse Clustering in BOSS}\label{boss2}
As described above, the angle-averaged clustering 
(the monopole) constrains the combination $D_{\mathrm{A}}^2/H(z)$.
The transverse signal probes $D_{\mathrm{A}}(z)$, 
which in turn is related to an integral of $H(z)$. 
This is limits  the constraining power of $D_{\mathrm{A}}(z)$ 
due to degeneracies. 
The line-of-sight feature, however, constrains 
$H(z)$ directly at the mean sample redshift. Therefore, 
by measuring at the feature in various $z$ slices, 
we can measure its change over time.

In previous sections we demonstrated that a 
narrow line-of-sight signal is not obtainable in 
the SDSS-II LRG sample, and should be noisy in 
BOSS.

\begin{figure*}[htp]
\epsscale{2.4}
\plotone{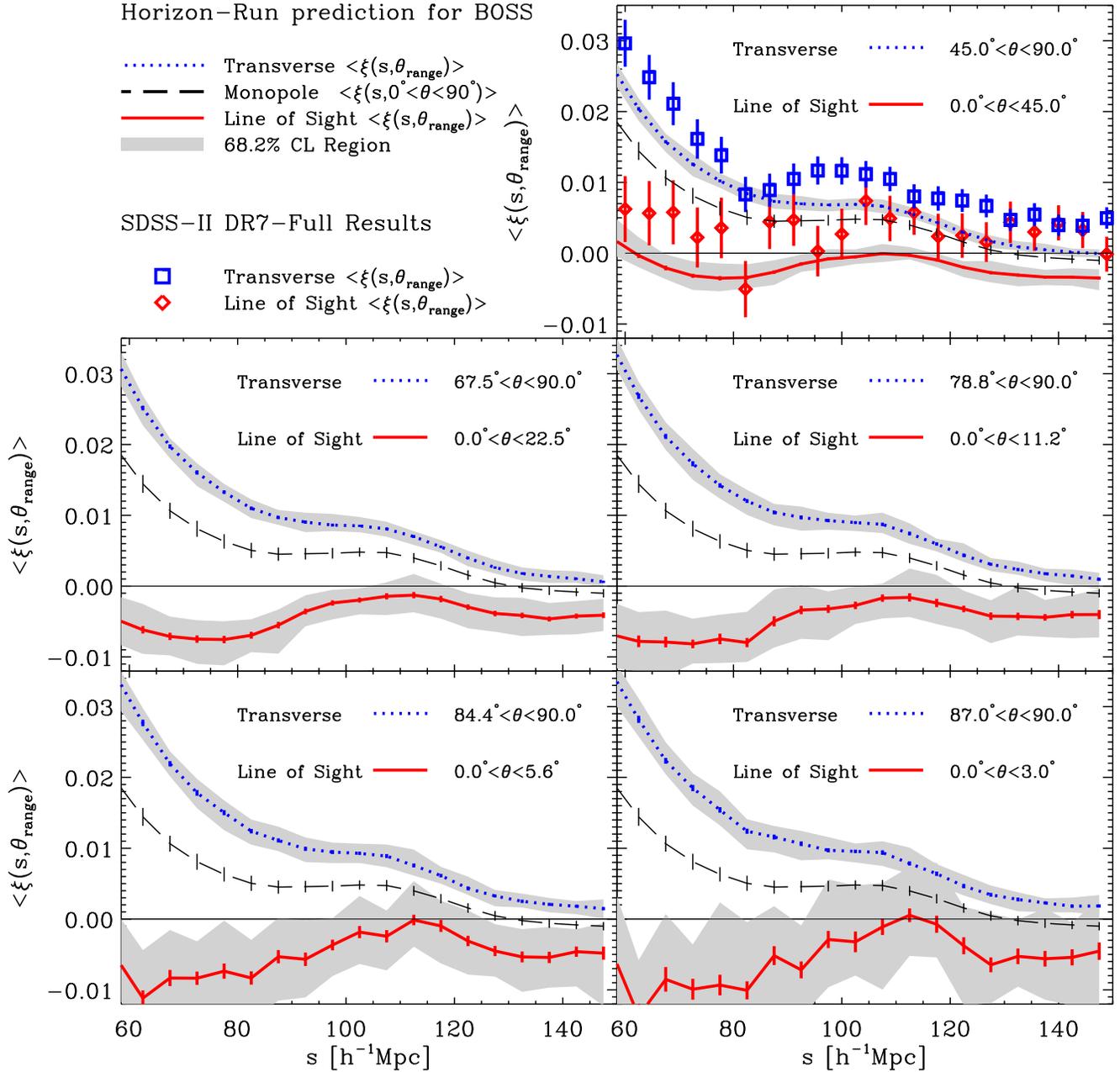}
\caption{BOSS expected line-of-sight and transverse angular slices of $\xi$. 
{\small 
Each panel shows two angular wedges of the redshift-space correlation
function $\xi(s)$ as predicted for BOSS from the Horizon Run
simulations. The size of the wedges are indicated in each panel
(starting at $45^\circ$ at the top and decreasing towards the
bottom). The solid red lines show the wedges closest to the
line-of-sight direction, and the dotted blue lines show the wedges
closest to the tranverse direction. In each case we give
the 1$\sigma$ uncertainty in the mock mean as the error bars and the
uncertainty for a single BOSS volume as the gray band. The angular
slices are $\Delta\theta= 45^\circ,\ 22.5^\circ,\ 11.2^\circ,\
5.6^\circ $ and $3^\circ$. The dashed line is the monopole prediction
and is the same in all panels, with uncertainties given for one BOSS
volume (i.e., not the mock mean). The symbols in the top right panel
are the result for DR7-Full in the $\Delta\theta= 45^\circ$ slices,
where red diamonds are for the line-of-sight direction and the blue
squares are for the transverse direction.
}
}
\label{anglecut}
\end{figure*}

A higher signal-to-noise ratio can be obtained 
in the data by 
using larger angular slices. 
In Figure \ref{anglecut} we follow the 
same procedure as before: this time with 
wider angular slices. 
The top right panel displays results for 
angular slices split at $\theta=45^\circ$. 
The solid red line is the line-of-sight result, now defined as 
$\left<\xi(s,0^\circ<\theta<45^\circ)\right>$ 
and the dotted blue line is the transverse result 
$\xi(s,45^\circ<\theta<90^\circ)$. 
DR7-Full results are the symbols. 
The expected 
redshift-space signals are the thick lines; the 
gray bands are the $1\sigma$ uncertainties expected from one BOSS volume. 
For comparison, the expected monopole 
($\left<\xi(s,0^\circ<\theta<90^\circ)\right>$)
is the dashed black line. 
The uncertainties in the transverse signal are 
smaller than the line-of-sight, because the former  contains 
much more information than the latter
(c.f. the $\sin(\theta)$ factor in Equation \ref{xisa}).

The clear vertical offset between the line-of-sight signal 
and the transverse signal
is a result of redshift 
distortions. 
We note that 
in the resulting region where the correlation
function is negative, there is a clear sign 
for a trough-peak-trough. 

As a consistency check for our software, 
when analyzing the Horizon Run real-space 
catalog, 
we have verified that the offset in signal 
is not present; in real-space, both the
line-of-sight and transverse correlation functions
have a similar signal as the monopole, as expected.

In the subsequent panels we slice the anisotropic $\xi$ into finer
angular divisions (each $\Delta\theta$ is indicated in the legend and
the caption). Again for each wedge size, we display the wedges closest
to the line-of-sight direction (solid red line), and closest to the
transverse direction (dotted blue line). We plot the monopole result in each
panel for comparison (black dashed line).
Although the signal is noisier for smaller wedges, 
a clear detection of the relative peak position in each direction 
appears obtainable by BOSS. 

We defer to the future a thorough analysis of the precision with which
BOSS will measure $H(z)$ and $D_{\mathrm{A}}$ by measuring the 
\baf as a function of angle $\theta$. 
Here we simply note that the low signal-to-noise ratio
expected for the very narrow line-of-sight cone studied by \Gpaper
should not discourage us from this measurement.

This independent measure of the 
line-of-sight and transverse features in clustering 
can be interpreted as a test of the
\cite{alcock79} effect.
When counting pairs we have assumed a fiducial 
cosmology for the purpose of converting the 
observed redshifts to comoving distances. 
Although this cosmology is well motivated 
by WMAP 5-year results (\citealt{komatsu09a}), small deviations 
from the true underlying cosmology will result 
in distortions in $\xi$, and of most concern, 
the position of the \bafii. 
 
\cite{alcock79} describe how an intrinsically 
spherical body in real-space appears distorted 
to an observer who measures the object in redshift-space and 
uses an incorrect cosmology to convert to comoving space. 
In real-space the \baf appears as such a spherical 
body (in the statistical sense) and the line of 
sight \baf should yield the same result as the 
transverse direction. 

The spherical nature of the \baf is only approximate in redshift-space 
because it
is somewhat distorted due to gravitational dynamics.  However, the
effect of these dynamic distortions on the position of the feature are
understood. Thus, by comparing the line-of-sight feature to the
transverse feature, we can learn about the true underlying cosmology.

We emphasize that in Figure \ref{anglecut} we merely  
show disentanglement of line-of-sight and transverse clustering signals. 
We do not perform the \cite{alcock79} test as the fiducial 
cosmology used to calculate mock $\xi$ is the same as the simulations.

\section{Discussion and Conclusions}\label{conclusions}

In this paper we demonstrate that the claim of a 
{\it
significant detection
}
in the line-of-sight \baf in the SDSS LRG sample by \Gpaper is
unjustified. 
We perform a similar analysis to theirs, 
and obtain similar results. The main 
difference in our interpretation, as we elaborate in \S\ref{interpretation},
is that we  use a more conservative criterion 
regarding whether we detect a feature. 
We find that the data agrees very well with a $\Lambda$CDM 
redshift-space non-linear model tested here, which 
does not contain a
clear 
 line-of-sight feature due to its
 low signal-to-noise ratio. 
 We also find that physical line-of-sight models 
 tested by us and 
 \Gpaper 
 do not  out-perform 
 a null $\xi=0$ model,  
 indicating 
 no clear evidence of a line-of-sight \bafii. 
  The BOSS survey, which has just begun, 
will have the statistical power
to rule out (or confirm) this strong clustering excess at high significance
(Figure \ref{horizon_run}),   
though not to usefully
detect the \baf in such a narrowly defined line-of-sight
measurement. By using broader angular bins, BOSS will be able to
independently measure \baf along the line-of-sight and transverse
directions.

We examine two different volumes in 
the SDSS LRG sample (SDSS-II). 
In the smaller one ($0.16<z<0.36$; \qvlii; \S \ref{dr7dim_analysis}),  
for which we use very realistic mock catalogs,  
we find a good agreement ($1\sigma-1.5\sigma$) 
between the 
line-of-sight observation and a $\Lambda$CDM model, 
which does not have an apparent 
feature even in $\sim 160$ \qvl volumes.  
For our full sample ($0.16<z<0.47$; DR7-Full) 
we find that a fiducial model agrees 
within $1.5\sigma-2\sigma$. 

Figures \ref{LOS_SDSS_LD}, \ref{los_dr7full_plot}, \ref{sn} 
clearly show that the line of
sight clustering excess at $s\sim 110$\hmpc 
in both volumes 
is dominated by 
noise.

We confirm the number of pairs estimate of 
\cite{miralda09} which they used on \Gpaper results.  
The bin of interest is 
$106.7<s<111.1$\hmpc and $0^\circ<\theta<3^\circ$, corresponding to
the choice in \Gpaper of $\Delta \pi$ and $r_p<5.5$\hmpcii.  In our
case, we count the number of effective pairs, including the effects of
weighting (see \Kpaperalt \ for details). 
\qvl yields $2259$ pairs and DR7-Full yields $3104$. 
Examining $160$ very realistic mock catalogs, 
we find that the high \qvl value of $\xi=0.085$ in this bin, 
although unlikely, is consistent with $\Lambda$CDM
(one mock of $160$ has a higher value). 

We point out that the strong clustering excess 
that 
{\it
happens to be at the correct scale 
} 
is 
{\it
probably 
}
noise, that will be reduced with increase of volume. 
We see a hint of this effect
 when the volume increases from 
\qvl ($0.66$\hgpcii) to DR7-Full ($1.6$\hgpcii) 
the line-of-sight clustering excess ($\theta<3^\circ$) at  $s \sim 108.9$\hmpc is decreased from 
$\xi \sim 0.085$ to $\xi \sim 0.045$.
This is shown also in \Gpaper in their Figures 13 and 15.  We
interpret the difference between both measurements and the expected
value from $\Lambda$CDM as noise.

Assuming correctness of our fiducial cosmology,  
the predicted BOSS results, presented in 
Figure \ref{horizon_run}, 
show that the strong line-of-sight 
clustering excess seen in DR7 should be ruled out at 
a very high confidence level when 
measuring $\xi$ in 
the BOSS LRG sample. 

In the left panels of Figure \ref{sn} 
we show the low S/N expected from 
the line-of-sight measurement in 
BOSS. 
We also examine by eye all $32$ line-of-sight results 
(defined as $r_p<5.5$\hmpcii), 
and find that $7$ of $32 \  (\sim 22\%)$ mocks do 
not show a clear line-of-sight \bafii, where $12$ do show 
a clear signal $(\sim 38 \%)$. 
This analysis is, of course, subjective, 
but suggests that the LRGs alone, 
might not detect the line-of-sight 
\baf in the BOSS volume,  
if the line-of-sight definition is limited to $\theta<3^\circ$ 
or $r_p<5.5$\hmpcii.

In Figure \ref{anglecut} we show, however, 
reason for optimism in 
disentangling $H(z)$ and $D_{\mathrm{A}}(z)$
in the future 
BOSS LRG data. 
Unsurprisingly, using wider angular cuts yields 
better signal-to-noise ratio results.
The various $\Delta\theta$  
slices contain line-of-sight and 
transverse information that 
may be used to probe $H(z)$ and 
$D_\mathrm{A}(z)$. 

Throughout this study, we have ignored the effects of any potential
reconstruction techniques that might be employed on the BOSS data
(\citealt{eisenstein07}, \citealt{huff07a}, \citealt{padmanabhan09a}, \citealt{noh09a}).   
Reconstruction may allow better precision in measuring
the peak position than indicated here.

We emphasize that our measurements 
and uncertainty estimates are similar 
to those carried out previously by \Gpaperii. 
We also confirm that the strong 
clustering signal at the line-of-sight 
can not be explained by systematics. 
We check various binning widths in 
both $\pi$ and $r_p$ as well 
as fiber-collision weightings, 
and do not find significant changes. 
We do, however, disagree on interpretation 
of the results.

The main dispute is regarding 
the use of the strong line-of-sight clustering  in the data 
as the \baf to determine H(z) directly. 
In their Table 3 \Gpaper show results for 
$H(z)$ obtained by two different methods: 
\begin{enumerate}
\item ``Shape Method": 
using the full shape of the line-of-sight clustering 
with priors from monopole, quadrupole clustering as well  as CMB temperature fluctuations  
to determine $H(z$)/$H_0$.
\item ``Peak Method": using the line-of-sight \baf peak position to determine $H(z)$ directly. 
\end{enumerate}
While both methods are correct procedures to perform, 
the second should be considered valid only 
if the line-of-sight \baf is convincingly detected.   

To explain the insignificance of a detection, we consider the null $\xi=0$ test. 
Using a Jeffreys scale, in \S\ref{interpretation} we show 
that the physical models do not 
out-perform $\xi=0$. 
This does not mean, of course, 
that $\xi=0$ is the line-of-sight clustering 
at these scales, but rather indicates 
that models can not be distinguished 
significantly and 
a clear line-of-sight \baf 
can not be declared detected 
using this data set. 
 

Only future surveys can show  
definitively what the line-of-sight correlation function
is in these bins.  In particular, BOSS will be able to
do so.
In Figure \ref{horizon_run} 
we show 
that BOSS, 
which is underway, 
will be able to rule out this detection 
or verify it. 
BOSS is 
due to cover a comoving volume of ($\sim8.1$\hgpcii; $0.16<z<0.80$) by 2014.
For a comoving volume of $\sim3.9$\hgpcii ($0.16<z<0.60$)  and density 
expected in BOSS, we find 
that all our $32$ mock realizations have an 
apparent {\it monopole} feature 
(as opposed to SDSS-II volumes, \Kpaperalt). 

Another survey that is on its way to refine \baf measurements 
is the WiggleZ. 
The WiggleZ Dark Energy Survey (\citealt{drinkwater10a}) is expected to complete very soon 
a narrower redshift survey (area $\sim 1,000$ deg$^2$), using Blue
Emission Line Galaxies, 
in a comoving volume of $\sim 1$Gpc$^3$ between $0.2<z<1$. 
When analyzing a quarter of the 
predicted final sample, 
\cite{blake10a} show in Figure $20$ a hint of a baryonic acoustic wiggle in the P$(k)$ monopole. 
Although it is not clear whether this survey will detect 
a significant peak in the line-of-sight $\xi$ 
(assuming $\theta<3^\circ$),
we are hopeful it will give us indication 
of the true signal.

If the claim for detection by \Gpaper is 
correct, the unexpected strong line-of-sight signal would require 
an explanation. 

It is a pleasure to thank Enrique Gazta\~{n}aga for 
commenting in detail on our draft,  in depth discussions  on 
interpretation, uncertainty estimation issues and sharing information 
regarding analysis. We thank Lam Hui and David Hogg 
for discussing interpretation of results, as well as Anna Cabre 
who shared with us her results.   
We thank Nic Ross for his very useful comments which helped clarify our draft. 
We thank Daniel Eisenstein for assistance in selecting LRGs and 
Idit Zehavi for discussions on weighting algorithms and selection function considerations. 
We thank Nikhil Padmanabhan, Nic Ross and David Schlegel for BOSS related discussions. 
We also thank Vincent Desjacques, Abraham Loeb, Ariel S$\acute{\mathrm{a}}$nchez, and Martin White 
for useful discussions and insight. 
We thank the LasDamas collaboration (http://lss.phy.vanderbilt.edu/lasdamas/) for making their mock catalogs publicly available. 
Thanks are also due to the Horizon team (http://astro.kias.re.kr/Horizon-Run/) for making their mocks public,
and in particular  Changbom Park and Juhan Kim for discussions on usage.
E.K thanks Jo Bovy and Mulin Ding for their technical help.
E.K thanks David Schlegel and Nic Ross at 
Lawrence Berkeley National Lab 
for their hospitality during part of the analysis in March.
E.K was partially supported by a Google Research Award and NASA Award
NNX09AC85G.  M.B was supported by Spitzer G05-AR-50443 and NASA Award
NNX09AC85G. R.S. was partially supported by NSF AST-0607747 and NASA NNG06GH21G.

 Funding for the SDSS and SDSS-II has been provided by the Alfred P. Sloan Foundation, the Participating Institutions, the National Science Foundation, the U.S. Department of Energy, the National Aeronautics and Space Administration, the Japanese Monbukagakusho, the Max Planck Society, and the Higher Education Funding Council for England. The SDSS Web Site is http://www.sdss.org/.

 The SDSS is managed by the Astrophysical Research Consortium for the Participating Institutions. The Participating Institutions are the American Museum of Natural History, Astrophysical Institute Potsdam, University of Basel, University of Cambridge, Case Western Reserve University, University of Chicago, Drexel University, Fermilab, the Institute for Advanced Study, the Japan Participation Group, Johns Hopkins University, the Joint Institute for Nuclear Astrophysics, the Kavli Institute for Particle Astrophysics and Cosmology, the Korean Scientist Group, the Chinese Academy of Sciences (LAMOST), Los Alamos National Laboratory, the Max-Planck-Institute for Astronomy (MPIA), the Max-Planck-Institute for Astrophysics (MPA), New Mexico State University, Ohio State University, University of Pittsburgh, University of Portsmouth, Princeton University, the United States Naval Observatory, and the University of Washington.




\begin{deluxetable}{ccccccccccc}
\rotate
\tabletypesize{\footnotesize}
\small
\tablewidth{551pt}
\tablecaption{SDSS-II and BOSS LRG Samples}{\label{tabledata}}
\tablehead{
\colhead{Sample} & 
\colhead{\# of LRGs} &
\colhead{$z_{\mathrm{min}}$} &
\colhead{$z_{\mathrm{max}}$} & 
\colhead{$\avg{z}$}  &
\colhead{$M_{g,\mathrm{min}}$} & 
\colhead{$M_{g,\mathrm{max}}$} &
\colhead{$\avg{M_g}$} & 
\colhead{Area} & 
\colhead{Volume} & 
\colhead{Density} \\
& 
&
&
& 
&
&
&
&
\colhead{(deg$^{2}$)} & 
\colhead{($h^{-3}$ Gpc$^{3}$)} & 
\colhead{($10^{-5}$ $h^3$ Mpc$^{-3}$)} \\
}
\startdata
 \qvl      &                                $61,899$  & $0.16$    &    $0.36$    &  $0.278$  & $-23.2$ & $-21.2$  &  $-21.65$ & $7,189$     & $0.66$     &     $9.4$    \\   
DR7-Full  &                           $105,831$ & $0.16$    &   $0.47$    &  $0.324$  & $-23.2$    & $-21.2$  &  $-21.72$ & $7,908$  & $1.58$     &   $6.7$    \\       
BOSS\tablenotemark{a} & $1,200,000$  & $0.16$  &  $0.60$     &   $0.444$  & $-$  & $-$  &  $-$ & $10,000$  & $3.89$    &   $30.8$  \\   
\enddata
\tablenotetext{a}{For BOSS we present estimates for $0.16<z<0.6$. 
The survey intends to observe $z<0.8$ ($8.1$\hgpcii) but the comoving number density is expected to fall sharply at $z>0.6$}

\end{deluxetable}

\begin{deluxetable}{ccccccccc}
\rotate
\tabletypesize{\footnotesize}
\small
\tablewidth{551pt}
\tablecaption{Jeffreys Scale Test for Line-of-Sight Models\label{jeffreys}}
\tablehead{
\colhead{Data} &
\colhead{Model} &
\colhead{$\chi^2$} & 
\colhead{$k$} &
\colhead{$N$} &
\colhead{$\chi^2/(N-k)$} &
\colhead{$Prob [\%]$\tablenotemark{c} } &
\colhead{$-\ln E \sim \chi^2+kln(N)$} &
\colhead{$-\ln E \sim \chi^2+2k$} \\
}
\startdata
 \Gpaper DR6  ($0.15<z<0.30$)& No BAO              &            $25$ &  $5$    &   $20$   &$1.7$ &  $5$& $40.0$     &  $35.0$ \\  
  \Gpaper DR6 ($0.15<z<0.30$) & BAO              &            $21$ &  $5$    &   $20$   &$1.4$ & $13$ &$36.0$     &  $31.0$ \\       
  \Gpaper DR6 ($0.15<z<0.30$) & BAO$+$mag             & $15$  &   $6$  &     $20$  &$1.1$   & $38$  & $33.0$    &  $27.0$\\   \\

  \Gpaper DR6  ($0.15<z<0.47$)& No BAO              &            $18$ &  $5$    &   $20$   &$1.2$ &  $26$ &$33.0$    &  $28.0$ \\  
  \Gpaper DR6  ($0.15<z<0.47$)& BAO              &            $12$ &  $5$    &   $20$   &$0.8$ &  $68$ &$27.0$     &  $22.0$ \\       
  \Gpaper DR6  ($0.15<z<0.47$) & BAO$+$mag             & $8$  &   $6$  &     $20$    &$0.6$ &  $89$ &$26.0$    &  $20.0$\\   \\

 \Gpaper  DR6 ($0.15<z<0.47$)&  $\xi=0$                &      $14$  &  $0$    &   $20$  &$0.7$  &  $83$ &$14.0$   & $14.0$    \\ 
    \Gpaper DR6  ($0.15<z<0.47$)& BAO              &            $12.6$ &  $0$    &   $20$   &$0.6$ &  $89$ &$12.6$     &  $12.6$ \\    
 \Gpaper DR6  ($0.15<z<0.47$) & BAO$+$mag             & $9.7$  &   $1$  &     $20$    &$0.5$ &  $96$ &$12.7$    &  $11.7$\\   
   \Gpaper  DR6 ($0.15<z<0.47$)\tablenotemark{d} &  $\xi=0$                  &      $15.8$  &  $0$    &   $20$  &$0.8$  &  $73$ &$15.8$   & $15.8$    \\  
   \Gpaper DR6  ($0.15<z<0.47$)\tablenotemark{d} & Horizon Run              &            $17.2$ &  $0$    &   $20$   &$0.9$ &  $64$ &$17.2$     &  $17.2$ \\   \\    
 
  \qvl    ($0.16<z<0.36$)            & $\xi=0$                          &     $23.2$  & $0$ &     $20$  &$1.2$ &$28$ &  $23.2$ & $23.2$ \\
  \qvl     ($0.16<z<0.36$)          & LasDamas &  $24.8$  & $0$ &     $20$   & $1.2$ &$21$&  $24.8$    & $24.8$ \\ \\
  
  DR7-Full   ($0.16<z<0.47$)\tablenotemark{b}            & $\xi=0$  &     $31.8$  & $0$ &     $20$  & $1.6$  & $4.5$& $31.8$ &  $31.8$ \\
  DR7-Full ($0.16<z<0.47$)\tablenotemark{b}       & Horizon Run&  $28.7$  & $0$ &     $20$  & $1.4$    &$9.4$&$28.7$ & $28.7$\\ 
    DR7-Full   ($0.16<z<0.47$)\tablenotemark{b}            & $\xi=0$ &     $13.7$  & $0$ &     $10$  & $1.4$  & $19$& $13.7$ &  $13.7$ \\
  DR7-Full ($0.16<z<0.47$)\tablenotemark{b}       & Horizon Run&     $12.6$  & $0$ &     $10$  & $1.3$    &$25$&$12.6$ & $12.6$ \\ \\
  
  
\enddata
\tablenotetext{b}{For DR7-Full we use the $C_{ij}$ based on 160 \qvl mocks normalized by the $C_{ii}$ of 32 DR7-Full mocks}
\tablenotetext{c}{Assuming a $\chi^2$ distribution this answers: What is the probability to obtain a random variable from a $\chi^2$ distribution larger than in the $\chi^2$ column?}
\tablenotetext{d}{Here we apply our own DR7-Full $C_{ij}$\tablenotemark{b} to the DR6 data in Table 1 of \Gpaperii. To the $C_{ij}$ we also apply a normalization of the ratio of volumes of DR7 and DR6, $1.58/1.35$, which takes into account different sky coverage as well as the fact that \Gpaper use LRGs at $z>0.15$. }
\tablecomments{{\large $\xi_{\mathrm{Line-of-Sight}}$ Model Statistics:} {\normalsize Models are as mentioned in text. The $\xi=0$, although not physical, gives indication of how noisy the data is. The $\chi^2$ is a statistic that tests how well each model fits the data. To compare goodness of two models calculate $\Delta\ln E$ (we present both BIC and AIC values). Following Table 1 from \cite{liddle09a}:  $\Delta \ln E<1$ "Not worth more than a bare mention",  $1<\Delta \ln E<2.5$ "Significant",  $2.5<\Delta \ln E<5$ "Strong to very strong",  $5<\Delta \ln E$ "Decisive". For descriptions of LasDamas and Horizon Run mocks, please refer to \S\ref{mocks}. All tests are conducted on bins in the range $\pi=[40,140]$\hmpcii.}} 

\end{deluxetable}

\end{document}